\documentclass{aa} 
\usepackage{graphicx}
\usepackage{txfonts} 
\begin{document}
 \title{Dust properties in \object{M31}.I.}

 \subtitle{Basic properties and a discussion on age-dependent dust heating}

 \author{M. Montalto \inst{1,2}, S. Seitz \inst{1,2}, A. Riffeser \inst{1}, U. Hopp \inst{1,2}, C.-H. Lee \inst{1}, R. Sch$\rm \ddot o$nrich \inst{1}}

 \offprints{}

 \institute{
            University Observatory Munich, Scheinerstrasse 1, 81679 Munich, Germany
            \email{montalto@usm.uni-muenchen.de}
 \and
            Max-Planck Institute for Extraterrestrial Physics, Giessenbachstrasse, 85748 Garching, Germany\\
 }
 
 \date{\today}

\newcommand{\M}{\mathrm}
\newcommand{\U}{\,\mathrm}
\newcommand{\FUV}{\mathrm{FUV}}
\newcommand{\NUV}{\mathrm{NUV}}
\newcommand{\TIR}{\mathrm{TIR}}

 
 \abstract
 {
 }
 { 
 Recent observations derived from the Spitzer Space Telescope and
 improvements in theoretical modeling of dust emission properties
 are used to discuss the distribution of dust and its
 characteristics in the closest neighbor spiral galaxy
 \object{M31}. Together with GALEX FUV, NUV, and SDSS images we
 studied the age dependence of the dust heating process. 
 }
 {

 Spitzer IRAC/MIPS maps of \object{M31} were matched together and
 compared to dust emission models allowing to constrain the dust
 mass, the intensity of the mean radiation field, the abundance
 of Polycyclic Aromatic Hydrocarbons (PAH)
 particles. The total infrared emission (TIR) was analyzed in
 function of UV and Optical colors and compared to predictions of
 models which consider the age-dependent dust heating.
 }
 {We demonstrate that cold-dust component emission dominates the
 infrared spectral energy distribution of \object{M31}. The mean
 intensity of the radiation field heating the dust is low
 (typically $U<2$, where $U=1$ is the value in the solar
 surrounding). Due to the lack of submillimetric measurements the
 dust mass ($M_\M{dust}$) is only weakly constrained by the
 infrared spectrum, but we derived a lower limit of $M_\M{dust}
 \gtrsim 1.1 \times 10^{7}M_{\odot}$ with a best fit value of
 $M_\M{dust} = 7.6\times 10^7\,M_{\odot}$, in good agreement with
 expectations from CO and HI measurements. We show that across
 the spiral-ring structure of \object{M31} a fraction $>3\%$ 
 of the total dust mass is in PAHs. UV and optical colors are correlated to
 total infrared to far ultraviolet 
 ($\TIR/\FUV$) ratios in $\sim 670\U{pc}$-sized regions overall
 the disk of \object{M31}, although deviating from the relationship between
 infrared excess and ultraviolet spectral slope
 (referred as $\M{IRX}$-$\beta$ relationship) for starburst 
 galaxies. In particular, redder regions show lower values of the $\TIR/\FUV$ ratio 
 for a fixed color. Considering the predictions of models that account for the
 dust-heating age dependence
 we derived that in $83\%$ of the regions analyzed
 across the 10kpc ring more than $50\%$ of the energy absorbed by 
 the dust is rediated at $\lambda > 4000 \AA$ and that
 dust in \object{M31} appears mainly heated by populations a few Gyr old 
 even across the star-forming ring. We also found that the attenuation is varying radially peaking
 near 10kpc and decreasing faster in the inner regions of \object{M31}
 than in the outer regions in agreement with previous studies. We finally 
 derived the attenuation map of \object{M31}
 at 6$\rm ^{''}/px$ resolution ($\sim 100\U{pc/px}$ along the plane of \object{M31}).
 }

 \keywords{\object{M31} -- dust -- extinction
 }

 \maketitle
%

\section{Introduction}
\label{s:introduction}

The presence of the dust affects astrophysical observations in
different ways. While in the UV-optical spectral region the dust
absorbs the stellar radiation field, it re-emits the absorbed energy 
in the IR-FIR and sub-millimetric spectral regions. It is thus 
essential to determine the properties of the dust in a given system  
as accurately as possible. While in our own galaxy we
can reach very high spatial resolutions and accuracies, our particular
location inside the galaxy limits our analysis.

Being the closest large spiral like the Milky Way, \object{M31} 
offers          a          fascinating        view       of   the dust 
distribution        across       the disk of the galaxy and an unique
opportunity to study the properties of a major disc galaxy in
detail. Previous studies already focused on the dust in this galaxy,
but new observational results and theoretical improvements require
some issues to be revisited and critically discussed and allow to shed
new light onto still unresolved problems. In particular GALEX UV
observations (Gil de Paz et al.~2007), SDSS optical images (York et
al. 2000), and Spitzer Space Telescope observations (Barmby et
al. ~2006, Gordon et al. ~2006) in the infrared and far infrared are now
available for \object{M31}. Together with the recent CO and HI
observations (Nieten et al.~2006, Braun et al.~2009) this new wealth
of data allows a much more detailed exploration of the dust properties
out than in previous studies. Especially Spitzer data allow to study
significantly smaller scales than in the IRAS era.
\\
Moreover, in the last years important progress has been made on the
theoretical understanding of dust properties facilitating the
derivation of new parameters from the data (Weingartner \& Draine
2001; Draine \& Li 2007; Draine et al. 2007). This work is thus
devoted to an analysis of recent data available for \object{M31} in
light of the new theoretical instruments. We focus our attention on
the mean radiation field heating the dust, on the dust mass
determination and on the amount and distribution of Polycyclic
Aromatic Hydocarbon particles (PAHs) in \object{M31}.

\object{M31} offers a nice example for studying a problem that has
drawn a lot of attention in recent studies: what is the main source
heating the dust in a galaxy? Certainly the dust tends to
preferantially absorb UV photons, so naively one would of course
expect UV radiation especially from the young populations seen in the
GALEX images of \object{M31} to be the main source powering the
observed dust emission. Indeed this idea clearly holds for
starburst and very actively star forming galaxies (see e.g. Buat 1992,
Meurer et al.~1995) and triggered the discovery of the IRX-$\beta$
relation (Meurer et al. 1999), but in galaxies hosting older stellar
populations the radiation field is clearly dominated by longer
wavelengths. Recent studies have demonstrated that in systems with low
relative star formation rate old stars can contribute significantly to
the dust heating (Cortese et al.~ 2008; Kong et al.~2004, Buat et
al.~2005; Gordon et al.~ 2000) and we will see in this paper that the
latter applies for \object{M31}. This difference is crucial if one
attempts to derive accurate dust attenutations and consequently star
formation rates in any galactic system.

The structure of this paper is as follows: In Sect.~\ref{sec:data} we
present the examined data and the calibration procedure, in
Sect.~\ref{sec:models} we describe the dust models and their
application to determine the dust mass, the mean intensity of the
radiation field responsible for dust heating and of the PAHs
abundances in \object{M31}. In Sec.~\ref{sec:pop} we discuss the age
dependency of dust heating, deriving the $A_\FUV$ attenuation across
the spiral-ring structure of \object{M31}. Finally in
Sects.\ref{sec:discussion} and \ref{sec:conclusions} we discuss and
summarize our results.

\section{The data}
\label{sec:data}

\subsection{IR}

In this work we used the IR images of \object{M31} described in Barmby
et al.~(2006) and Gordon et al.~(2006), and obtained with the
IRAC/MIPS instruments on board of the {\it Spitzer Space Telescope}
(SST). In particular, we retrieved from the SST archive the $3.6\U{\mu
  m}$, $8\U{\mu m}$ IRAC, and the $24\U{\mu m}$, $70\U{\mu m}$ and
$160\U{\mu m}$ MIPS Basic Calibrated Data (BCDs) which were processed
by the SST's archive pipeline (v.14). The BCDs were then stacked
together using the MOPEX software (version 18.1).

For IRAC observations we applied overlap correction before mosaicing
in order to match the background between adjacent images and discarded
the first frame of each observing run because it is typically acquired
with a lower bias level than subsequent images (first frame effect).
The background was evaluated in the external regions and subtracted
from the images with the IRAF {\it imsurfit} task using a low order
polynomial (first or second order depending on the map). We then
applied the IRAC photometric corrections for infinite sources as
reported in the IRAC documentation multiplying the $3.6\U{\mu m}$
image by $0.91$ and the $8\U{\mu m}$ image by $0.74$ in order to
account for scattering of incident light in the array focal planes. We
applied the correction for extended sources because in this work we
focus on the diffuse dust emission, not on point-like sources.

For MIPS $70\U{\mu m}$ and $160\U{\mu m}$ data, we used our customized
software for background subtraction via fitting and subtracting a
first order polynomial along the scan direction excluding the
\object{M31} region.  The final MIPS mosaic images have a resolution
of $6\arcsec/\M{px}$, $18\arcsec/\M{px}$ and $40\arcsec/\M{px}$ at
$24\U{\mu m}$, $70\U{\mu m}$ and $160\U{\mu m}$, respectively.

Because our purpose was to directly compare the fluxes in each band on
a pixel by pixel basis we had to resample the images to the same
astrometric reference system. In fact we created two sets of images:
in the first one the IRAC $3.6\U{\mu m}$ and $8\U{\mu m}$ images were
put to the reference system of the $24\U{\mu m}$ mosaic
($6\arcsec/\M{px}$), in the second one the IRAC $3.6\U{\mu m}$ and
$8\U{\mu m}$ and the MIPS $24\U{\mu m}$ and $70\U{\mu m}$ were put on
the reference system of the $160\U{\mu m}$ mosaic
($40\arcsec/\M{px}$). These two sets will be considered separately in
the following. To make the different instrument images comparable, we
first matched the PSF of the input instrument to that one of the
target instrument (MIPS $24\U{\mu m}$ or MIPS $160\U{\mu m}$). This
was done with the help of the transformation kernels provided by
K. Gordon\footnote{http://dirty.as.arizona.edu/~kgordon/mips/conv\_psfs/conv\_psfp.html}
(Gordon et al.~2008).
These kernels give the transformation between the PSFs of each couple
of IRAC-IRAC, IRAC-MIPS and MIPS-MIPS instruments dependent on the
dust temperature. After the convolution we resampled the images to the
same astrometric grid, in such a way that each pixel in a given band
corresponded to the same physical region of the same pixel in the
other bands. We estimated the uncertainty of the resampling step by
means of a comparison of the integrated photometry on the whole galaxy
before and after this procedure and found that in the worst case the
relative variation in flux was $3\%$, as shown in
Tab.~\ref{tab:percvar}. The uncertainty in the background subtraction
was estimated evaluating the scatter in the background subtracted
mosaics in the outermost regions and found to be negligible.

\begin{table}
  \caption{Percentual variation of the total integrated flux of
    \object{M31} as measured before and after the resampling process of
    the images described in the text.  The third column presents the
    percentual variation for the images resampled to the $24\U{\mu m}$
    resolution, the third column for the images resampled to $160\U{\mu
      m}$. The total integrated flux was measured in the same area in
    all the images, correspondent to an ellipse centered on
    \object{M31}, with a semi-major axis of $84\arcmin$, a semi-minor
    axis of $18\arcmin$, and a position angle of $\M{PA} =
    38^{\circ}$.}
  \label{tab:percvar} 
  \centering 
  \begin{tabular}{c c c} 
    \hline\hline 
    Band ($\M{\mu m}$) & resampled to $24\U{\mu m}$ ($\%$) & resampled to $160\U{\mu m}$ ($\%$) \\ 
    \hline 
    3.6 & 2 & 2   \\ 
    7.9 & 3 & 2   \\
    24  & - & 0.8 \\
    70  & - & 0.6 \\
    160 & - & -   \\
    \hline 
  \end{tabular}
\end{table}

While creating the mosaics with {\it MOPEX}
\footnote{http://ssc.spitzer.caltech.edu/postbcd/mopex.html}
(MOsaicker and Point source EXtractor), we derived also the associated
error map, which provides the error related to the whole reduction
procedure performed by the pipeline. To account for the other sources
of error (resampling, background subtraction, BCDs calibration) we
simply added a constant term equal to $10\%$, with except of the
$24\U{\mu m}$ map for which we would have otherwise obtained
unrealistically large values of the error. In reality our approach can
be considered rather conservative, because as stated in the SST
documentation the standard deviation maps account for the error of the
pipeline in a conservative way. This is also demonstrated by a
comparison of our own measurements of the integrated fluxes of
\object{M31} in each band with those derived by Barmby et al.~(2006)
and Gordon et al.~(2006) as shown in Tab.~\ref{tab:integrated}. In
particular, with respect to the errors quoted by Barmby et al.~(2006)
and Gordon et al.~(2006), our integrated flux measurements are
0.6$\sigma$ smaller, 0.1$\sigma$ smaller, $1\sigma$ larger,
$0.8\sigma$ larger and $0.4\sigma$ smaller than their own measurements
at $3.6\U{\mu m}$, $8\U{\mu m}$, $24\U{\mu m}$ $70\U{\mu m}$ and
$160\U{\mu m}$ respectively, and our estimated error is $9\%$ smaller,
$30\%$, $54\%$, $36\%$ and $50\%$ larger than their given
uncertainties. We conclude that our measurements are consistent within
$1 \sigma$ and that our errors are in general equivalent or larger
with respect to the values derived by Barmby et al.~(2006) and Gordon
et al.~(2006).

\begin{table*}[h!]
  \begin{minipage}[t]{\textwidth}
    \caption{Measurements of \object{M31} infrared integrated fluxes
      obtained using IRAC (Barmby et al. 2006), MIPS (Gordon et
      al. 2006), COBE (Odenwald et al. 1998), IRAS (Rice et al. 1988),
      MSX (Kraemer et al. 2002) and ISO (Haas et al. 1998)
      instruments.  The last column shows our own measurements derived
      from IRAC and MIPS observations.  We measured the integrated
      flux in an ellipse of semi-major axis $84\arcmin$ and semi-minor
      axis $18\arcmin$ centered on \object{M31} with a  position
        angle of $\M{PA} = 38^{\circ}$.}
    \label{tab:integrated} 
    \centering 
    \renewcommand{\footnoterule}{} 
    \begin{tabular}{l l l c c c c c c c} 
      \hline\hline 
      $\lambda$\footnote{The wavelengths reported in the first column are the nominal
                         wavelengths of each instrument waveband. The minimum and maximum
                         wavelengths reported in the second and third columns correspond to a
                         system response (filter+detector) equal to 10$\%$ the maximum. In
                         particular the system response functions were taken for Spitzer/IRAC
                         from http://ssc.spitzer.caltech.edu/irac/spectral\_response.html,
                         for Spitzer/MIPS from
                         http://ssc.spitzer.caltech.edu/mips/MIPSfiltsumm.txt, for COBE/DIRBE
                         from
                         http://lambda.gsfc.nasa.gov/data/cobe/dirbe/ancil/spec\_resp/DIRBE\\
                         \_SYSTEM\_SPECTRAL\_RESPONSE\_TABLE.ASC, for IRAS from
                         http://irsa.ipac.caltech.edu/IRASdocs/exp.sup/ch2/tabC5.html, for
                         MSX/SpiritIII from
                         http://irsa.ipac.caltech.edu/data/MSX/docs/MSX\_psc\_es.pdf
                         (Appendix A, Table A-3), for ISO/ISOPHOT from
                         http://www.mpia.de/ISO/welcome.html.  }
      & $\lambda_\M{min}^{a}$ & $\lambda_{max}^{a}$ & This work & IRAC & MIPS & COBE & IRAS & MSX & ISO \\ 
      $\M{\mu m}$ & $\M{\mu m}$ & $\M{\mu m}$ & Jy & Jy & Jy & Jy & Jy & Jy & Jy \\
      \hline 
         1.27 & 1.1   & 1.5   &                 &             &                 & 534 $\pm$ 107   &                &            &                \\ 
         2.22 & 2.0   & 2.4   &                 &             &                 & 461 $\pm$ 92    &                &            &                \\ 
         3.53 & 3.0   & 4.1   &                 &             &                 & 245 $\pm$ 49    &                &            &                \\ 
         3.55 & 3.2   & 3.9   & 239$\pm$29      & 259$\pm$32\footnote{Uncertainties in this table have been directly
                                                  taken from the literature when available. For COBE measurements
                                                  we assumed an error equal to 20$\%$ of the measured flux in each band 
                                                  as suggested by Odenwald et al.~(1998) and for IRAS measurements 
                                                  equal to 15$\%$ as suggested by Rice et al.~(1988).
                                                  For our own measurements the uncertainties have been derived as detailed
                                                  in the text}&                 &                 &                &            &                \\ 
         4.88 & 4.5   & 5.3   &                 &             &                 & 128 $\pm$ 26    &                &            &                \\ 
         4.49 & 4.0   & 5.0   &                 & 144$\pm$ 20 &                 &                 &                &            &                \\
         5.73 & 5.0   & 6.5   &                 & 190$\pm$35  &                 &                 &                &            &                \\
         7.87 & 6.4   & 9.5   & 149$\pm$27      & 151$\pm$ 21 &                 &                 &                &            &                \\
          8.3 & 6.0   & 10.9  &                 &             &                 &                 &                & 159$\pm$32 &                \\
         12.0 & 7.6   & 15.4  &                 &             &                 &                 & 163$\pm$ 24    &            &                \\ 
         23.7 & 20.5  & 28.5  & 118$\pm$17      &             & 107 $\pm$ 11    &                 &                &            &                \\
         25.0 & 16.5  & 30.4  &                 &             &                 &                 & 108$\pm$ 16    &            &                \\ 
         56.0 & 38.6  & 76.5  &                 &             &                 & 700 $\pm$ 140   &                &            &                \\ 
         60.0 & 37.0  & 82.8  &                 &             &                 &                 & 536 $\pm$ 80   &            &                \\ 
         71.4 & 55.1  & 91.6  & 1086 $\pm$ 256  &             & 940 $\pm$ 188   &                 &                &            &                \\
         97.7 & 68.6  & 121.8 &                 &             &                 & 3706 $\pm$ 741  &                &            &                \\ 
        100.0 & 74.4  & 130.6 &                 &             &                 &                 & 2928 $\pm$ 439 &            &                \\ 
        155.9 & 128.9 & 184.0 & 7315 $\pm$ 1632 &             & 7900 $\pm$ 1580 &                 &                &            &                \\ 
        147.9 & 108.2 & 181.7 &                 &             &                 & 7545 $\pm$ 1509 &                &            &                \\ 
        175.0 & 140.0 & 220.0 &                 &             &                 &                 &                &            & 7900 $\pm$ 800 \\
        247.9 & 174.7 & 335.9 &                 &             &                 & 6242 $\pm$ 1248 &                &            &                \\ 
        \hline 
      \end{tabular}
   \end{minipage}
\end{table*}

We finally subtracted from the $8\U{\mu m}$ and the $24\U{\mu m}$ maps
the stellar continuum which was assumed to be described by the
$3.6\U{\mu m}$ map following Helou et al.~(2004):

\begin{equation}
\label{eq:helou1}
  F_{\nu}^\M{ns}(8\U{\mu m}) = F_{\nu}(8\U{\mu m}) - 0.232 F_{\nu}(3.6\U{\mu m})
\end{equation}

\begin{equation}
\label{eq:helou2}
  F_{\nu}^\M{ns}(24\U{\mu m}) = F_{\nu}(24\U{\mu m}) - 0.032 F_{\nu}(3.6\U{\mu m})
\end{equation}

\subsection{Ultraviolet}

The UV maps of \object{M31} were retrieved directly from the Nasa
Extragalactic Database (NED) and were obtained as part of the survey
of nearby galaxies performed with the {\it GALaxy Evolution eXplorer}
(GALEX) satellite (Gil de Paz et al.~2007). We retrieved and used both
the far ultraviolet (FUV) and the near ultraviolet (NUV) images.
These maps have a pixel scale of $1.5\arcsec/\M{px}$, a total
integration time of $800\U{sec}$ and are given in units of counts per
pixel per second (CPS). The background in these images can be assumed
constant and was accurately measured by Gil de Paz et al.~(2007) who
give a mean value of $(9.29 \pm 5.24)\times 10^{-4}$ CPS and $4.712
\pm 1.529) \times 10^{-3}$ CPS for the FUV and NUV respectively. We
subtracted the background from the maps and then converted the CPS
units in Jy multiplying by the factors $f_{0,\FUV}=108\U{\mu
  Jy}/\M{CPS}$ and $f_{0,\M{NUV}}=36\U{\mu Jy}/\M{CPS}$, as given by
the GALEX documentation. We then accounted for the foreground galactic
extinction assuming $E(B-V)=0.062$ in the direction of \object{M31} as
reported by Schlegel et. al.~(1998) with $A_\FUV = 7.9\,E(B-V)$ and
$A_\M{NUV} = 8.0\,E(B-V)$ as given by Gil de Paz et al.~(2007). Two
series of maps were finally derived in order to compare UV
observations with MIPS $24\U{\mu m}$ and MIPS $160\U{\mu m}$
data. Both the FUV and the NUV maps were convolved to the
correspondent MIPS PSFs and regridded at $6\arcsec/\M{px}$ and
$40\arcsec/\M{px}$. The global uncertainty of the final maps was
estimated to be $\sim\,10\%$ in flux for both maps.

\subsection{Optical}

We retrieved and stacked together the SDSS images of \object{M31}
obtained in the SDSS survey.  In this work we made use only of the
SDSS $i$ band (hereafter $i_\M{SDSS}$) images of \object{M31}.  First
we subtracted the SOFTBIAS in every single frame. SOFTBIAS was
originally added automatically by the SDSS pipeline to avoid negative
pixel values. We aligned each single frame to a large reference image
($1.5 \times 4.2$ square degree) using the IRAF task $wregister$. Then
we reconstructed each stripe. The final mosaic image contains 7
stripes (3366-2, 3367-3, 3366-3, 3367-4, 3366-4, 3367-5, and 3366-5,
from left to right in the mosaic image).  We first tested for
  photometric variation, but resulted in no zero-point difference
  between the stripes. Nevertheless we found sky variations in
  y-direction in different stripes when combining them into the mosaic
  image. Therefore, we applied an alignment in y-direction for each
  stripe. Using our customized written software ({\it mupipe/skycalc})
  we calculated the gradient $a\times y$ and the offset $b$ in flux of
  the overlapping areas between two adjacent stripes,
  e.g. $F_{3366-2}=F_{3367-3} + a_{3367-3} \times y+b_{3367-3}$ and
  applied the y-dependent sky correction to the whole stripe.  The
map was convolved with the Spitzer $160\U{\mu m}$ PSF and rescaled at
$40 \arcsec/\M{px}$ resolution.

Then we calibrated the images using cataloged surrounding stars with
SDSS photometry and masked out all the sources with $i<17\U{mag}$ as
most of the point-like sources in this magnitude range are foreground 
contaminants and excluding the brightest \object{M31} objects (globular
clusters, red supergiants, etc.) does not affect our analysis which
is focused on regions dominated by the dust diffuse emission.

\begin{figure}[!]
    \centering
    \includegraphics[width=9cm]{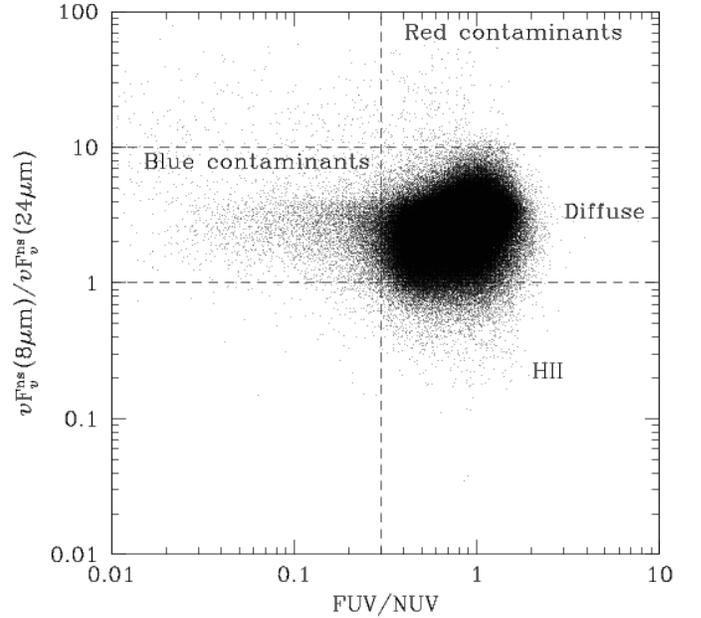}
    \caption{Two color diagram used to isolate the regions with dust
      diffuse emission.  Labels and dashed lines indicate different
      sources of emission in function of the colors. $(S/N)>1$ in all
      maps and each region corresponds to 6$\arcsec$/px resolution.  }
    \label{fig:twocolor}
  \end{figure}

  In order to select a un-contaminated sample of regions from the
  infrared and UV maps we used the two color diagram shown in
  Fig.~\ref{fig:twocolor}. We corrected the $8\U{\mu m}$ and $24\U{\mu m}$ 
  maps for the stellar continuum using Eq.~1 and Eq.~2 and worked at
  6$\rm ^{''}/px$ resolution. We limited the analysis to
  regions with $(S/N)>1$ by this excluding from the analysis the inner
  bulge region ($<3\arcmin$ from the center of \object{M31}),
  partially the interarm regions between the inner spiral-arm and the
  outermost regions of the galaxy. While the bulk of the points have
  colors with $1< \nu\,F^\M{ns}_{\nu}(8\mu
  m)/\nu\,F^\M{ns}_{\nu}(24\mu m)<10$ and $0.3<\FUV/NUV<2$, and appear
  after visual inspection associated with the diffuse dust emission,
  some deviant points are visible in Fig.~\ref{fig:twocolor}. In our
  maps bright point-like sources in the 24$\U{\mu m}$ map have
  $\nu\,F^\M{ns}_{\nu}(8\mu m)/\nu\,F^\M{ns}_{\nu}(24\mu m)<1$, thus
  lower than the ratios found in the diffuse component of the
  dust. Different studies have demonstrated that the emission from
  point-like sources at 24$\U{\mu m}$ is associated with the emission
  from HII regions in the UV and Optical spectral regions (Calzetti et
  al.~2007; Prescott et al.~2007).  Moreover, the PAHs emission at
  7.7$\U{\mu m}$ is expected to be strongly reduced with respect to
  the diffuse dust emission in such environments (e.g. Calzetti et
  al.~2005; Thilker et al.~2007; Bendo et al.~2008). On the other hand
  regions with $\nu\,F^\M{ns}_{\nu}(8\mu m)/\nu\,F^\M{ns}_{\nu}(24\mu
  m)>10$ or $\FUV/NUV<0.3$ appear to be associated with bright
  foreground stars. We thus adopt the selection criteria for regions:
  $1<\nu\,F^\M{ns}_{\nu}(8\mu m)/\nu\,F^\M{ns}_{\nu}(24\mu m)<10$ and
  $\FUV/NUV>0.3$. Once using also the SDSS optical map we further added
  the selection for excluding bright foreground stars as described above.

\begin{figure*}[!]
  \centering
  \includegraphics[width=15cm]{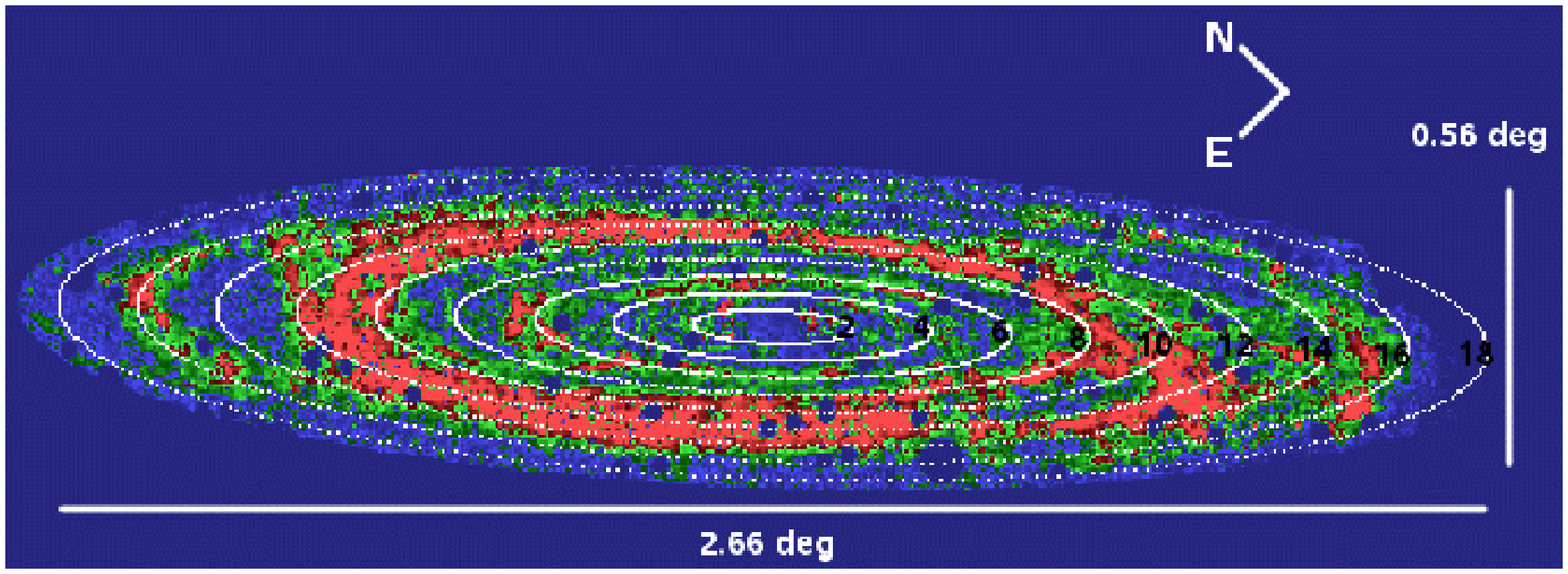}
  \includegraphics[width=15cm]{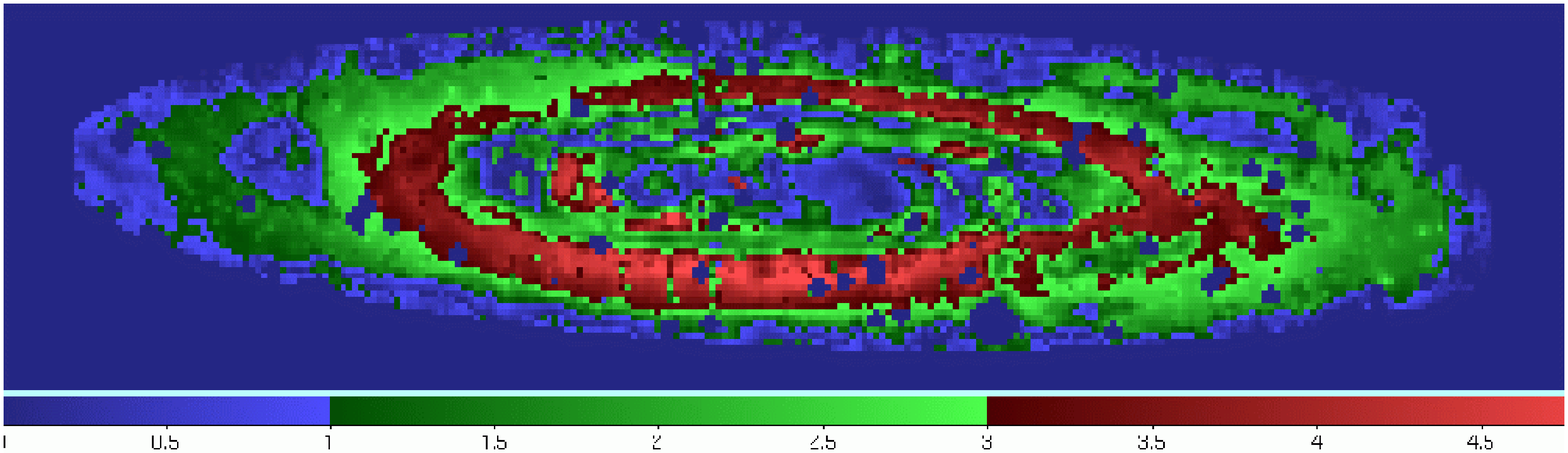}
  \caption{
    {\it Upper panel:} Minimum S/N in each one of the analyzed
    regions of \object{M31} once considering the 
    FUV, NUV, $\rm i_{\U{SDSS}}$, 3.6$\mu$m, 8$\mu$m, 24$\mu$m, 70$\mu$m and 160$\mu$m
    observational maps
    described in Sec.~\ref{sec:data} all together.  The resolution is
    6$\rm ^{''}/px$. We show also the nine concentric 2kpc-wide regions used in Sec.~\ref{sec:pop}.
    {\it Lower panel:} the same as the upper panel but
    for the 40$\rm ^{''}/px$ resolution maps.
    }
  \label{fig:sn}
\end{figure*}

  In Fig.~\ref{fig:sn} for each of the analyzed regions of
  \object{M31} is shown the {\it minimum} signal to noise ratio
  $(S/N)$ when considering all the above mentioned observational maps
  together.  The $(S/N)$ is in general larger than $3$ along the
  $10\U{kpc}$ ring and partially in the the inner spiral structure. In
  the interarm and external regions the $(S/N)$ drops below 1.5. In
  the bulge region the S/N is low because of the small stellar
  corrected flux at $8\U{\mu m}$.  


\section{The theoretical models}
\label{sec:models}

In this work we used and tested the theoretical models of infrared
emission of dust grains heated by starlight developed by Weingartner
\& Draine (2001) and Li \& Draine (2001), recently updated by Draine
\& Li (2007). In the following we summarize their main characteristics
and results.  Dust grains are treated as a mixture of amorphous
silicate and carbonaceous grains, where the smallest carbonaceous
grains have the physical properties of the PAHs. Temperature
distribution functions for all particles are calculated and emission
spectra provided for dust heated by a single radiation field
intensity. The spectral energy distribution for this is given by
Mathis, Mezger, \& Panagia (1983) (MMP83) scaled by the dimensionless
factor $U$, where $U = 1$ corresponds to the mean intensity of the
radiation field in the local ISM of the solar neighborhood.  We tested
the seven Milky Way dust models proposed by Weingartner \& Draine
(2001), which have PAH abundances with 0.47\%, 1.12\%, 1.77\%, 2.50\%,
3.19\%, 3.90\% and 4.58\% of the total dust mass.  These models assume
a dust-to-gas ratio $M_\M{dust}/M_\M{H} \simeq 0.01$.  For further
details the reader is referred to the studies mentioned above. The
emission spectrum predicted by the model can be approximated as
\begin{equation}
  \begin{array}{l}
    F_{\nu,\M{model}}=\Omega_\M{star}\,B_{\nu}(T_\M{star}) \, + \, \\
    \, + \,\frac{M_\M{H}}{4\pi D^{2}}\left[(1-\gamma)\,j_{\nu}^{(0)}
    \,(\M{model},U_\M{min})+\gamma\,j_{\nu}(\M{model},U_\M{min},U_\M{max})\right]
  \end{array}  
\end{equation}
where the first term accounts for the residual infrared emission
coming from the stars and is given by the product of $\Omega_\M{star}$
(the solid angle subtended by the stars, and $B_{\nu}(T_\M{star})$,
the blackbody emissivity with a fixed temperature $T_\M{star} =
5000\U{K}$, which was found by Smith et al.~(2007) to provide a good
description of the stellar continuum for $\lambda>5\U{\mu m}$. $D$ is
the distance of \object{M31} (we assumed $778\U{kpc}$),
$j_{\nu}^{(0)}(\M{model},U_\M{min})$ is the dust power radiated per
unit frequency per H nucleon when the dust is exposed to a single
radiation field of intensity $U_\M{min}$. The factor $\gamma$ ($0 \leq
\gamma \leq 1$) is introduced to parametrize the dust heating effects
of a power-law distribution of starlight intensities, as described in
Draine \& Li (2007). Actually, $j_{\nu}(\M{model},U_\M{min},U_\M{max})$ 
gives the dust power radiated per H nucleon by dust exposed to a power 
law distribution of starlight intensities ($\propto U^{-2}$) comprised 
between $U_\M{min} \leq U \leq U_\M{max}$. The fraction $\gamma$ 
($0 \leq \gamma \leq 1$) of dust emission associated with this 
intensity field represents dust emission close to OB associations and/or
photodissociation regions, where the intensity $U > U_\M{min}$; in
such a way this approach allows to handle dust temperature variations,
in a smooth and convenient way.  Following Draine et al. (2007) we
fixed in our calculation $U_\M{max} = 10^{6}$.  Finally $M_\M{H}$ is
the total mass of hydrogen (providing that the models give the
emission per H nucleon). The mass of the dust, $M_\M{dust}$, can be
obtained from $M_\M{dust} = M_\M{H}
\left(\frac{M_\M{dust}}{M_\M{H}}\right)$, assuming the dust to gas
ratio of the models.
\\
Each spectrum is thus completely characterized by $5$ free parameters:
the solid angle subtended by the stars $\Omega_\M{star}$, the kind of
emission model (amount of PAHs), the minimum intensity $U_\M{min}$ of
the starlight radiation field, the fraction $\gamma$ of the total dust
mass heated by the power-law distribution of starlight intensities,
and the total dust (hydrogen) mass $M_\M{H}$. The best fit is obtained
minimizing the quantity:

\begin{equation}
  \chi^{2} = \frac{1}{N_b}\sum_{b}^{N_b} \frac{\left(F_{\nu,\M{obs},b} \, - 
      \left<F_{\nu,\M{model}}\right>\right)^{2}}{\sigma_{\M{obs},b}^{2}+\sigma_\M{model}^{2}}
\end{equation}
 
\begin{figure*}[!]
 \centering
 \includegraphics[width=15cm]{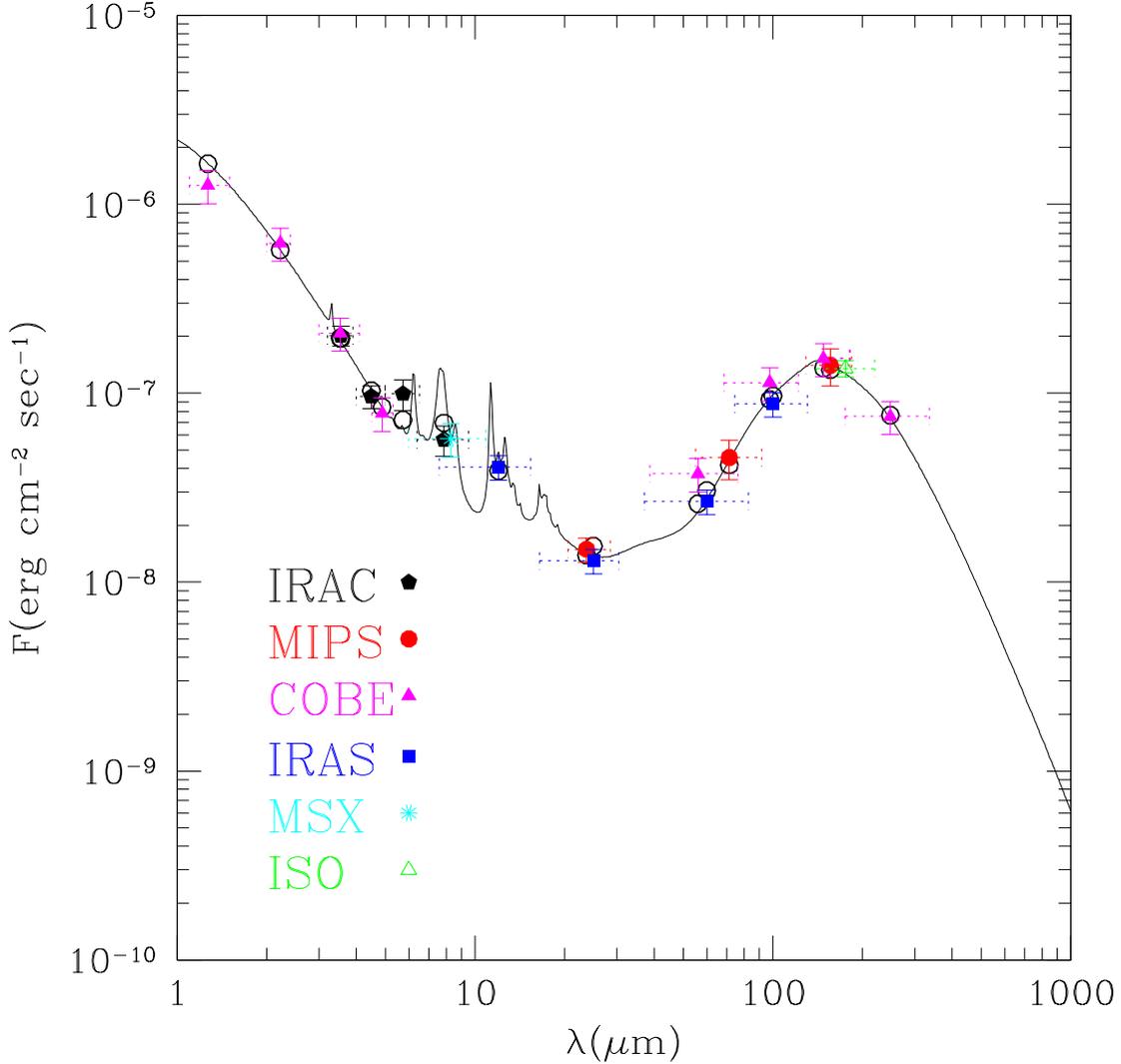}
 \caption{Infrared spectral energy distribution of \object{M31}. Color
   symbols are measurements obtained with the different instruments
   indicated in the legend.  The continuous line denotes the best
   fitting model of Draine et al.~(2007).  Open circles represent the
   model predicted fluxes after convolution with the instrumental
   response function, and these are the points that are compared with
   observations.  }
 \label{fig:spectrum_M31}
\end{figure*}

In Table~\ref{tab:integrated} we collected all the infrared integrated
flux measurements of~\object{M31} performed so far, obtained with
different instruments: IRAC (Barmby et al. 2006), MIPS (Gordon et
al. 2006), COBE (Odenwald et al. 1998), IRAS (Rice et al. 1988), MSX
(Kraemer et al. 2002) and ISO (Haas et al. 1998).  In the fourth
column we show our own measurements derived from IRAC and MIPS
observations.  In general all these measurements are in good
agreement, and define the spectral energy distribution of \object{M31}
from $\sim 1\U{\mu m}$ up to $\sim 250\U{\mu m}$, as shown in
Fig.~\ref{fig:spectrum_M31}.
\\
The best fit parameters we obtained were: $PAHs = 4.6\%$,
$U_\M{min} = 0.4$, $\gamma = 0$ and
$M_\M{dust}=7.6 \times 10^7\,M_\odot$, $\Omega_\M{star} = 3.1\times
10^{-16}\U{sr}$ and the $\chi^{2}$ of the best model was
$\chi^{2} = 0.55$.

In the following we discuss the uncertainties on the derived
parameters and the implications of the results.

\subsection{The uncertainty in the dust mass estimate}
\label{subsec:massestimate}

\begin{figure}[!]
  \centering
  \includegraphics[width=8cm]{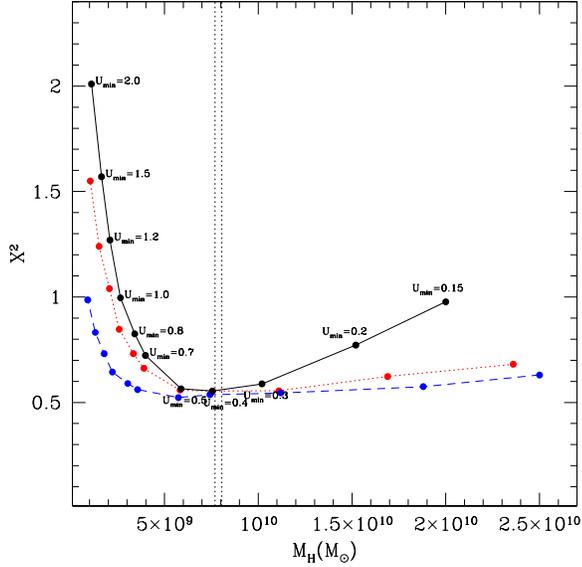}
  \caption{Minimum $\chi^{2}$ against $M_\M{H}$ derived from the model
    fits, for each fixed value of $U_\M{min}$ indicated by the
    labels. The black points (continuous line) represent the fitting
    results when considering all the measurements in
    Tab.~\ref{tab:integrated} (with the exclusion of ISO and MSX
    measurements), the red points (dotted line) when neglecting the
    COBE $248\U{\mu m}$ measurement, and the blue points (dashed line)
    when neglecting the MIPS $160\U{\mu m}$ and the COBE $148\U{\mu
      m}$, $248\U{\mu m}$ measurements.  The vertical dotted lines
    indicate the range of expected total hydrogen mass values
    estimated from HI and CO maps. Note that the points in this figure
    are connected by lines for clarity, but the $\chi^{2}$ curve is
    actually more complex, showing local minima for each value of
    $U_\M{min}$.}
  \label{fig:chi_mass}
\end{figure}

The mass estimate obtained in the above paragraph must be considered
rather uncertain.  This is due to the fact that, as shown in
Fig.~\ref{fig:spectrum_M31}, the peak wavelength of \object{M31}'s IR
spectrum is close to 160$\U{\mu m}$. Thus, the vast majority of our
analyzed observations is found at shorter waveleghts, therefore
colder mass components can't be reliably constrained by the present
observations. The only measurement that gives an appreciable
constraint is the COBE $248\U{\mu m}$ observation. Otherwise, the
cold-mass estimate critically depends on the temperature of the dust
inferred from the infrared spectrum.  For a given mean radiation field
intensity $U$ at which the dust is exposed the steady-state
temperature of the dust $T \propto U^{1/6}$ and the
emission peak $\lambda \propto U^{-1/6}$ (for grains with
dimension $>30\AA$, Draine \& Li 2007), that is $dT/T \propto
1/6\,dU/U$. By energy balance considerations $U$ is then directly
proportional to the emissivity per H nucleon of the dust grains
tabulated in the models and consequently the dust mass $M_\M{dust}$
derived comparing the model with the observations is $M_\M{dust}
\propto U^{-1}$, or $dM_\M{dust}/M_\M{dust} \propto -dU/U$. This makes
the dust temperature estimate relatively robust against uncertainties
in $U$, i.e. a variation of $100\%$ in $U$ (and in $M_\M{dust}$)
produces a variation of only $\sim 17\%$ in dust temperature and in
emission peak wavelength, to which our measurements are
sensitive. Moreover, $dU/U$ can be quite large even for small changes
of $U$ if $U$ is also small, as it seems to be the case for
\object{M31}. In order to perform a more quantitative analysis, in
Fig.~\ref{fig:chi_mass} we show the $\chi^{2}$ of the best model
obtained fixing the value of $U_\M{min}$, against the correspondent
total hydrogen mass $M_{H}$ predicted by the model.  The black points
(continuous line) in that figure show the result of the fit when all
measurements in Tab.~\ref{tab:integrated} are considered (with the
exclusion of ISO and MSX measurements which were not used here), the
red points (dotted line) when neglecting the COBE 248$\U{\mu m}$, and
the blue (dashed line) points when neglecting the MIPS 160$\U{\mu m}$
and the COBE 148$\U{\mu m}$ and 248$\U{\mu m}$ measurements. From that
is clear that the cold mass can't be accurately estimated. All the
curves show a shallow minimum in $\chi^{2}$ but the range of mass
values at which $\chi^{2}$ approaches the absolute minimum is large
even considering the fit with all the present observations. Moreover,
once FIR measurements are excluded the $\chi^{2}$ curves are even
flatter increasing the uncertainty in the mass values.  Nevertheless
it has to be noted that all $\chi^{2}$ curves are highly asymmetric
and imply that large mass values are much less constrained than low
mass as explained below.

\begin{figure}
  \centering
  \includegraphics[width=8cm]{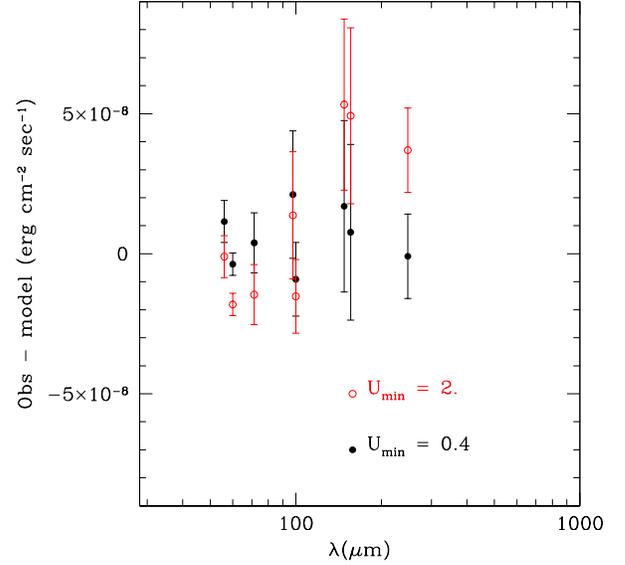}
  \caption{Observed minus model difference values in flux units for
    measurements with $\lambda > 50\U{\mu m}$.  All measurements
    reported in Tab.~\ref{tab:integrated} were used in the fit (with
    the exclusion of ISO and MSX measurements). Black filled points
    are the differences obtained considering the best fitting model
    ($U_\M{min} = 0.4$), whereas red open circles considering the best
    model among those ones with $U_\M{min} = 2$ fixed. Errorbars are
    the uncertainties associated with each measurement as given in
    Tab~\ref{tab:integrated}.}
  \label{fig:difference}
\end{figure}

In Fig.~\ref{fig:difference}, we present the difference between the
observations and the values predicted by two different models: (i) the
best fitting model (which has $U_\M{min}\,=\,0.4$, black filled
points); (ii) the best model obtained imposing $U_\M{min}\,=\,2$ (red
open circles).  All the measurements reported in
  Tab.~\ref{tab:integrated} are considered in the fit, thus the
$\chi^{2}$ of these two models are shown in Fig.~\ref{fig:chi_mass}
for the correspondent values of $U_\M{min}$ as given by the points
connected by the continuous black line.  In
  Fig.~\ref{fig:difference}, only measurements with
  $\lambda\,>\,50\U{\mu m}$ are considered because this is the
spectral region where the major differences can be noticed. A model
should be considered consistent with the observations if all the
observation minus model differences ($\Delta$) are consistent with
zero, within the observational errors. Focusing the attention on the
Far Infrared measurements (FIR) of Spitzer 160$\U{\mu m}$ and COBE
148$\U{\mu m}$, 248$\U{\mu m}$ in Fig.~\ref{fig:difference}, we
obtained for the best fitting model
$\Delta_{148\U{\mu m}} =  0.55\,\sigma_{148\U{\mu m}}$,
$\Delta_{160\U{\mu m}} =  0.24\,\sigma_{160\U{\mu m}}$ and
$\Delta_{248\U{\mu m}} = -0.06\,\sigma_{248\U{\mu m}}$,
whereas for the model with $U_\M{min} = 2$ we obtained
$\Delta_{148\U{\mu m}} = 1.74\,\sigma_{148\U{\mu m}}$,
$\Delta_{160\U{\mu m}} = 1.57\,\sigma_{160\U{\mu m}}$ and
$\Delta_{248\U{\mu m}} = 2.45\,\sigma_{248\U{\mu m}}$.
Thus in this spectral range the best fitting model with $U_\M{min}$
appears consistent with the observations at $<0.6\,\sigma$, whereas to
reconcile observations and model predictions once $U_\M{min} = 2$ we
have to admit differences $<2.5\,\sigma$, and in any case
$>1.5\,\sigma$. For smaller wavelengths the two models appear almost
equivalent in reproducing the observations, although a slightly worse
result is obtained for $U_\M{min} = 2$. An important feature of
Fig.~\ref{fig:difference} is that for $\lambda \gtrsim 140\U{\mu m}$
the flux model values for $U_\M{min} = 2$ are lower than the values of
the best fitting model, whereas for $\lambda \lesssim 140\U{\mu m}$
they are systematically larger. This is what we expect as the peak
emission of the model with $U_\M{min} = 2$ is shifted towards smaller
wavelengths with respect to the best fitting model, as a consequence
of the larger radiation field that is heating the dust. Because of
model predictions in the FIR for the case of $U_\M{min} = 2$ are
already at the limit of predicting an acceptable wavelength
distribution, larger radiation fields can be discarded. We conclude
that on the basis of the diffuse dust emission models of Draine \&
Li~(2007) and the analysis of the FIR spectrum of \object{M31} the
mean radiation field\footnote{In principle the value of $U_\M{min}$ is
  not coincident with the definition of the {\it mean radiation field}
  given by Draine \& Li (2007), although these two values are
  essentially equivalent once $\gamma \sim 0$ as seems the case for
  \object{M31}.} that is heating the dust in \object{M31} is typically
$U < 2$. As the radiation field is inversely proportional to the dust
mass estimate we can also conclude that the mass of the dust in
\object{M31} is $\gtrsim 1.1\times 10^{7}\,M_\odot$, which is the
value obtained for the  best model with $U_\M{min} = 2$, assuming
a dust-to-gas ratio equal to $1\%$.

The most recent and accurate estimate of the total mass of atomic
hydrogen (HI) in \object{M31} is that one of Braun et al.~(2009), and
is equal to $M_\M{HI}=7.33\times 10^{9}\,M_{\odot}$. Nieten et
al.~(2006) provided a mass of molecular hydrogen $M_{\M{H}_{2}}$ equal
to $M_{\M{H}_{2}}=3.6\times 10^{8}\,M_\odot$ (within a radius of
$18\U{kpc}$). These recent estimates combine to a total neutral hydrogen
mass of $M_\M{HI}+M_{\M{H}_{2}}=(7.69 \dots 8.05)\times 10^9\,M_\odot$
once considering an uncertainty range for the CO conversion factor
between $X_{HI}= (2\dots4) \times
10^{20}\U{mol}\U{cm}^{-2}\,(\M{K}\U{km}\U{sec}^{-1})^{-1}$.  The mass
estimate we obtained from the best fitting model (considering all the
observations in Tab.~\ref{tab:integrated}, $M_{H}=7.6\times
10^9\,M_{\odot}$) is thus respectively $1\%$ and $5\%$ smaller with
respect to the neutral hydrogen mass estimates at the extremes of the
uncertainty range mentioned above. Thus despite the large uncertainty
we dicussed previously, the best fitting model mass estimate appears
consistent with HI and CO measurements, as shown in
Fig.~\ref{fig:chi_mass}.  Earlier estimates of the dust mass
($M_\M{dust}$) in \object{M31} provided values equal to
$M_\M{dust}=3.8\times 10^{7}\,M_\odot$ (Haas et al. 1998) and
$M_\M{dust}=1.3\times 10^{7}\,M_\odot$ (Schmidtobreick et al. 2000)
which are fully consistent with our lower limit estimate but, for what
have been said so far, they cannot be considered more accurate than
our best model value estimate.

Draine et al.~(2007) analyzing a sample of 17 galaxies from the
SINGS-SCUBA sample have shown that in all these galaxies the value of
the mean radiation field derived from the best fitting model was $>2$,
with a median value of $4.3$. On that basis they proposed a restricted
fitting procedure for the case in which submillimetric observations
are not available which would imply to set $U_\M{min}>0.7$ during the
fit in order to avoid an overestimate of the dust mass.  Nevertheless
as stated by the same authors this procedure underestimates the dust
mass if the radiation field heating the dust is weak. The sample of
SINGS-SCUBA galaxies could be biased towards increased associated star
formation as recognized by the same authors.  As we have demonstrated
above, for the case of \object{M31} models with $U_\M{min}>2$ are not
in agreement with the FIR measurements. Moreover models with a strong
UV radiation field imply gas masses far below the range given by
neutral hydrogen mass measurements, whereas our best fitting model
($U_\M{min} = 0.4$) comes very close to the expected value. This
additional constraint rules out models with high $U_\M{min}$ and
proves that \object{M31} is a galaxy where the mean UV radiation field
heating the dust is very weak. In Sect.~\ref{sec:pop} we will provide
an independent demonstration of this fact and discuss its
implications.

\begin{figure*}
  \centering
  \includegraphics[width=8.5cm]{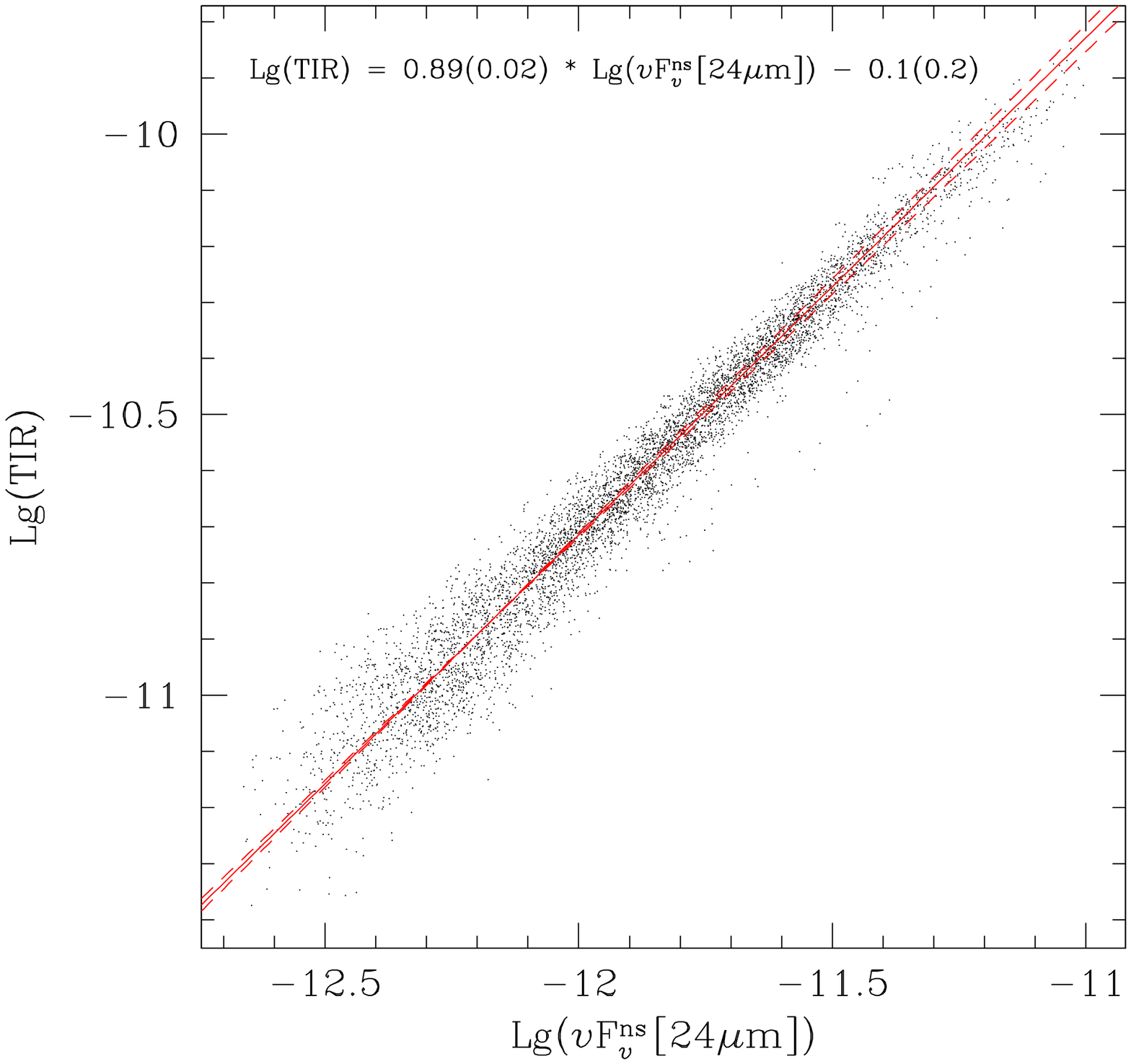}
  \includegraphics[width=8.5cm]{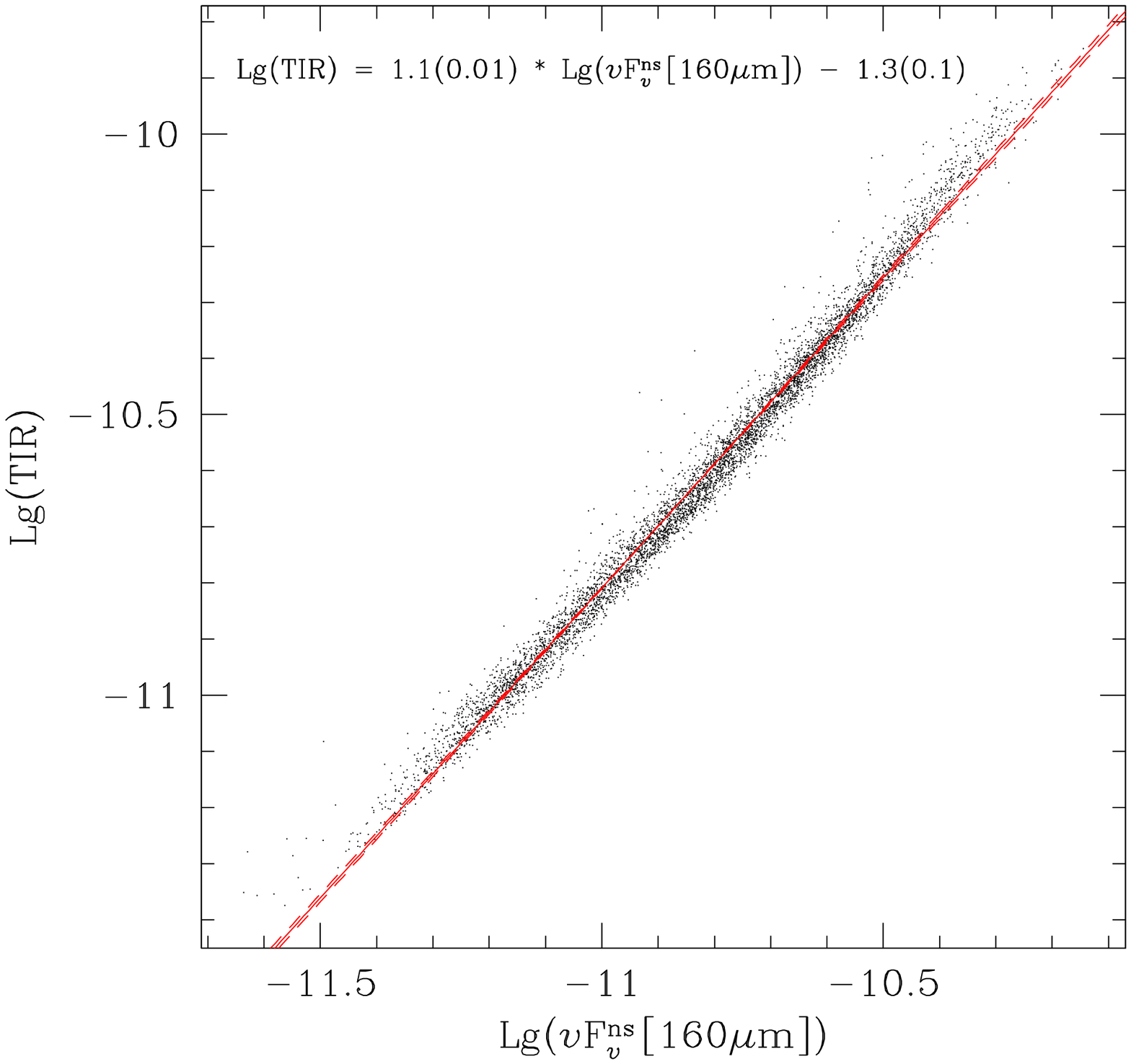}
  \includegraphics[width=8.5cm]{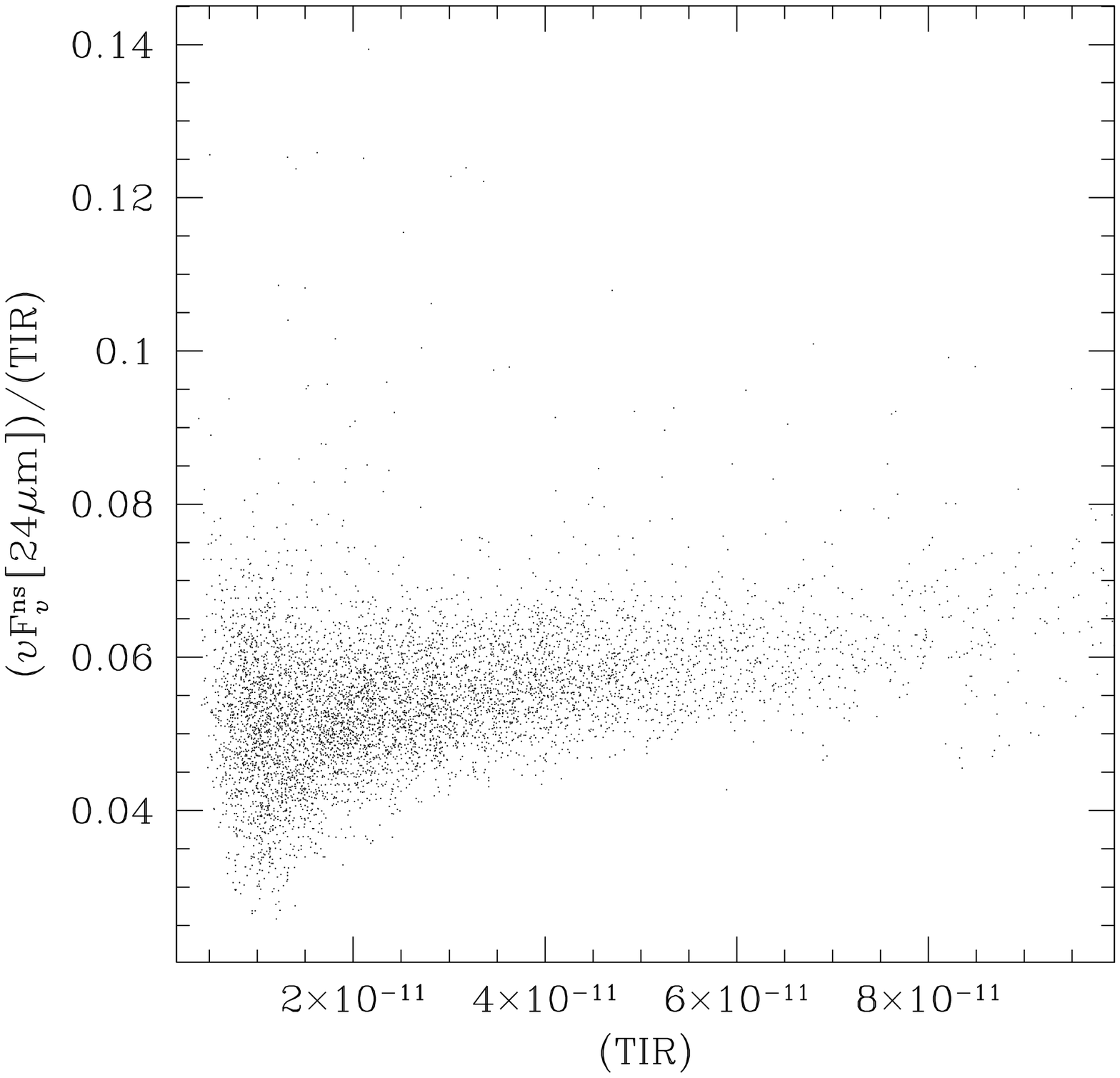}
  \includegraphics[width=8.5cm]{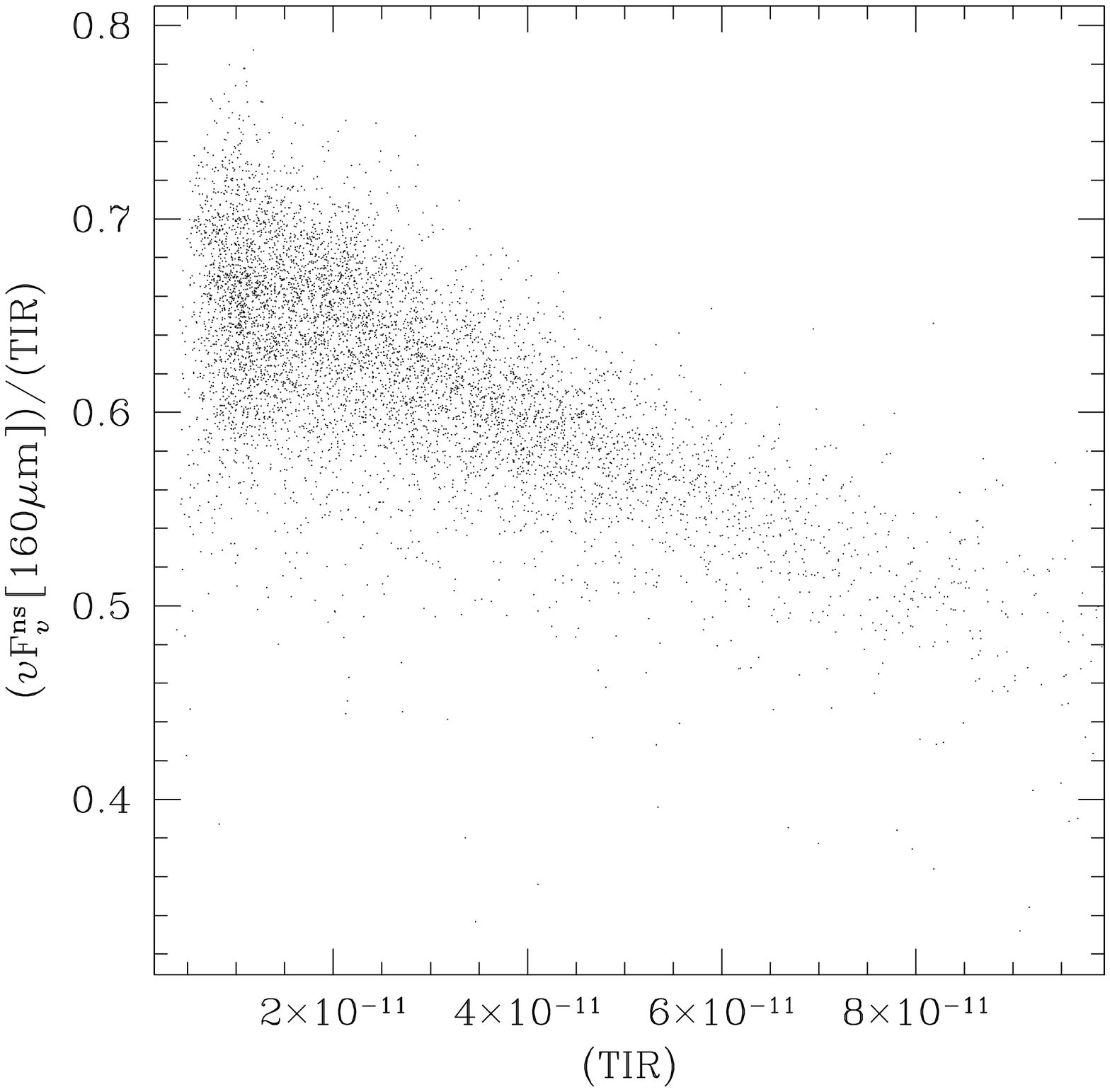}
  \caption{Upper panels: Correlations between the logarithm of the TIR emission and 
           the logarithm of the $24\U{\mu m}$ 
           ($160\U{\mu m}$) emission. Lower panels:
           correlations between the $24\U{\mu m}/TIR$ ($160\U{\mu m}/TIR$) 
           emission ratio and the TIR emission.                        
           Only regions with $\rm S/N>1$ are considered.
   }
  \label{fig:corr_tir}
\end{figure*}

In Fig.~\ref{fig:corr_tir} (upper panels) we show the correlation 
between the logarithm of the total IR (TIR) emission and the logarithm of the 
$24\U{\mu m}$ ($160\U{\mu m}$) emission. The TIR was 
estimated from the $Spitzer$ maps of
\object{M31} using the formula given by Draine \& Li~(2007):

\begin{equation}
   I_\TIR = 0.95\nu\,I_\nu(8\M{\mu m}) + 1.15\nu\,I_\nu(24\M{\mu m})
   + \nu\,I_\nu(70\M{\mu m}) + \nu\,I_\nu(160\M{\mu m})
\end{equation}

The $\U{lg}(160\U{\mu m})-\U{lg(TIR)}$ correlation is less scattered ($\M{RMS}=0.03$) 
with respect to the $\U{lg}(24\U{\mu m})-\U{lg(TIR)}$ correlation ($\M{RMS}=0.06$). 
We fitted these data with a simple linear model obtaining:

\begin{equation}
lg(TIR)=0.89(0.02)\,lg(\nu\,F_{\nu}^{ns}[24\mu\,m])-0.1(0.2)
\end{equation}

\begin{equation}
lg(TIR)=1.1(0.01)\,lg(\nu\,F_{\nu}^{ns}[160\mu\,m])-1.3(0.1)
\end{equation}

\noindent
Although, as shown below, the correlation between
the $24\U{\mu m}$ ($160\U{\mu m}$) and the TIR emission is not strictly linear,
these results (in particular the calibration relation between the $\U{lg(24\U{\mu m})}$ 
and the $\U{lg(TIR)}$ variables) will be useful later in Sect.~\ref{sec:discussion}
and sufficiently accurate for our purposes.

\noindent
In Fig.~\ref{fig:corr_tir} (bottom panels) we show the $24\U{\mu m}/\U{TIR}$
($160\U{\mu m}/\U{TIR}$) ratios against the TIR intensity. As the  TIR
increases the $24\U{\mu m}/\U{TIR}$ ratio increases, but remains confined
typically between $5\%$-$7\%$. Conversely, the $160\U{\mu m}/\U{TIR}$
decreases, changing from $\sim65\%$ to $\sim50\%$.
This demonstrates that overall the infrared spectrum of \object{M31} 
is dominated by the cold dust emission in the FIR, although the contribute
of hotter dust components tends to increase at larger TIR emissions.

\subsection{$PAHs$ abundance, $\gamma$}
\label{sect:param}

\begin{figure*}
  \centering
  \includegraphics[width=9.cm]{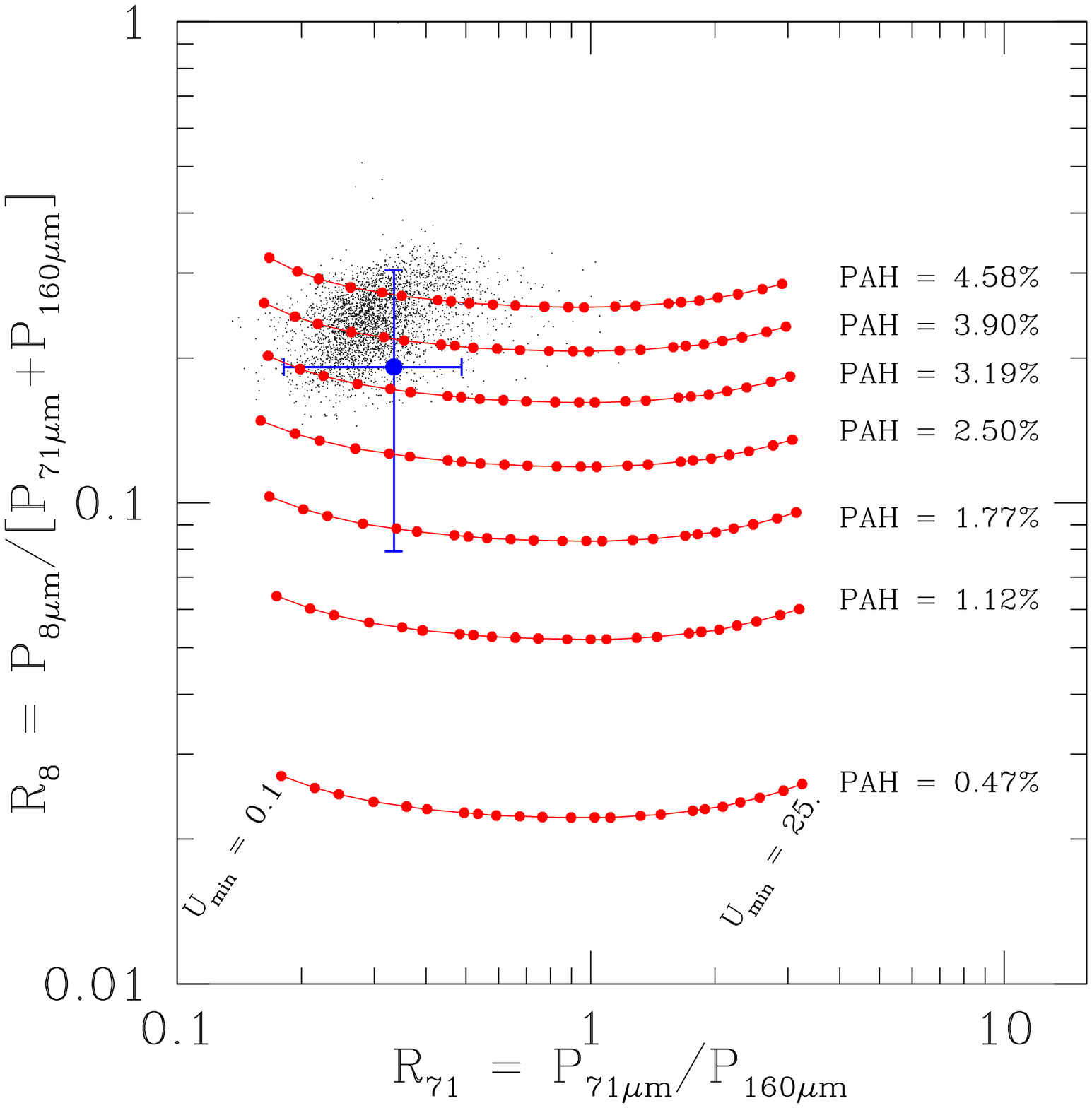}
  \caption{Ratio $R_8$ of the $8\U{\mu m}$ flux normalized to FIR
    emission against $71\U{\mu m}$ to $160\U{\mu m}$ flux
    $R_{71}$. The blue point is the ratio corresponding to the total
    integrated flux measurement of \object{M31}. Each horizontal red
    line shows the values predicted by the models of Draine \&
    Li~(2007) for different PAHs abundances and values of the minimum
    radiation field intensity $U_\M{min}$.}
  \label{fig:r8}
\end{figure*}

Fig.~\ref{fig:r8}, 
shows the ratio $R_8$ of the $8\U{\mu m}$ flux
(normalized to the total infrared emission) against the ratio $R_{71}$
of $71\U{\mu m}$ flux to $160\U{\mu m}$ flux. As proposed by Draine et
al.~(2007) this diagram can be used to estimate the abundance of
PAHs. We used it to illustrate the uncertainty of our fit comparing
the result obtained from the total integrated emission of \object{M31}
to that one obtained in the single regions of our maps, regridded at
the 160$\U{\mu m}$ resolution and using exclusively the regions with
$S/N>3$. Uncertainties were calculated as explained in
Sect.~\ref{sec:data}.  This selection restricted us to the study of
the ring-spiral structure.  Essentially PAHs are stochastically
heated by single photons thus in order to derive
their abundance one can compare the observations with theoretical
models with $\gamma = 0$  (Draine et al. 2007, where $\gamma$ is the parameter that
regulates the power-law intensity radiation field in Eq.~3). In
Fig.~\ref{fig:r8}, we show the theoretical predictions as functions of
the different amount of PAHs and of the minimum radiation field
intensity $U_\M{min}$.  The results imply that models with large
abundances ($>3.19\%$) and rather low $U_\M{min}$ are favoured when
comparing to the data. In fact the measurement associated with the
integrated infrared emission (blue point) is not at the center of the
distribution of points coming from the analysis of the pixel-to-pixel
analysis, and in particular is biased towards smaller values of $R_8$.
As our analysis is restricted to the spiral-ring structure, this
result suggests that in the inter-arm regions and/or toward the bulge
the abundance of PAHs could be smaller than what we derived, 
but an accurate analysis of the distribution of these particles 
across the whole disk of \object{M31} is beyond the 
purpose of this work.

\begin{figure*}
  \centering
  \includegraphics[width=9.cm]{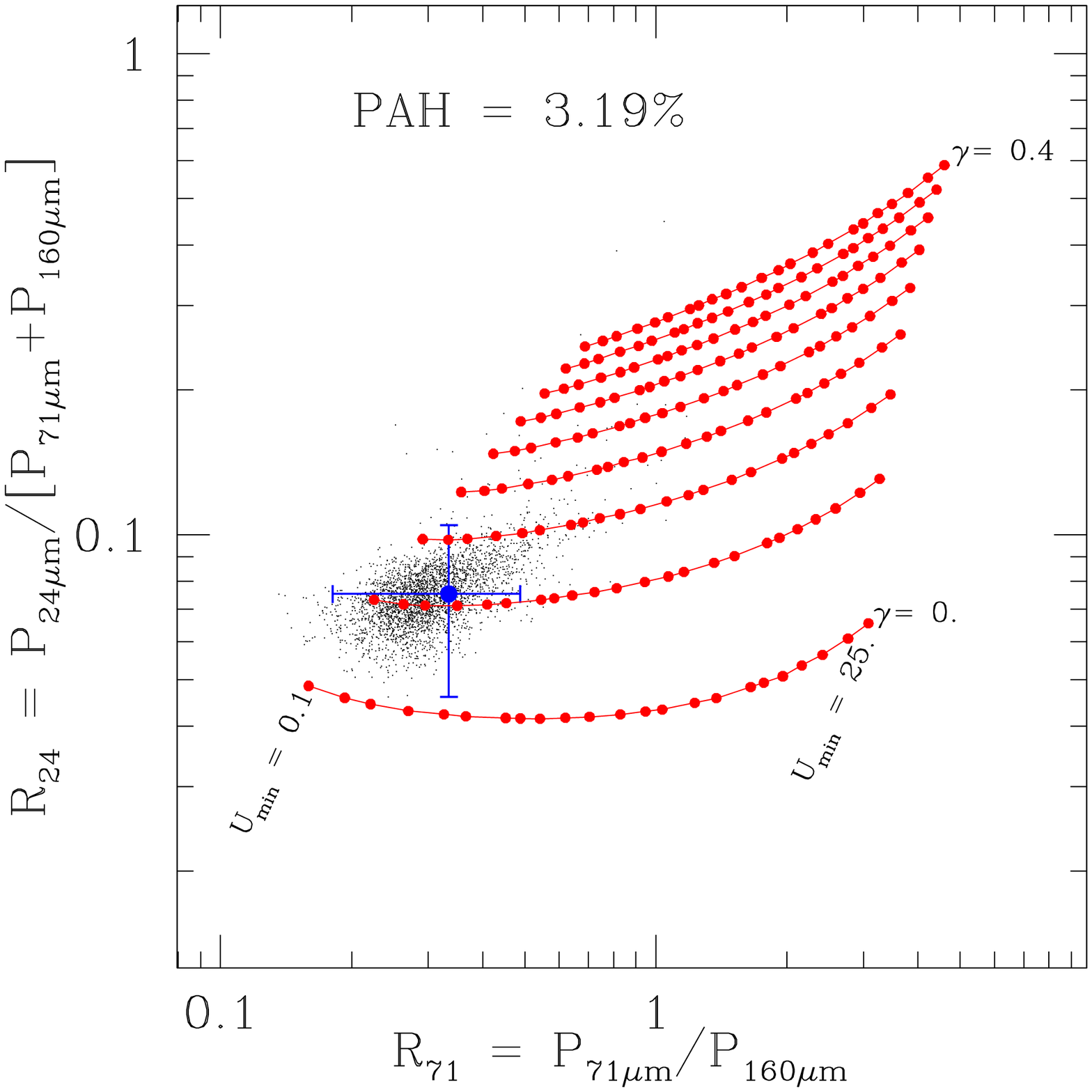}
  \includegraphics[width=9.cm]{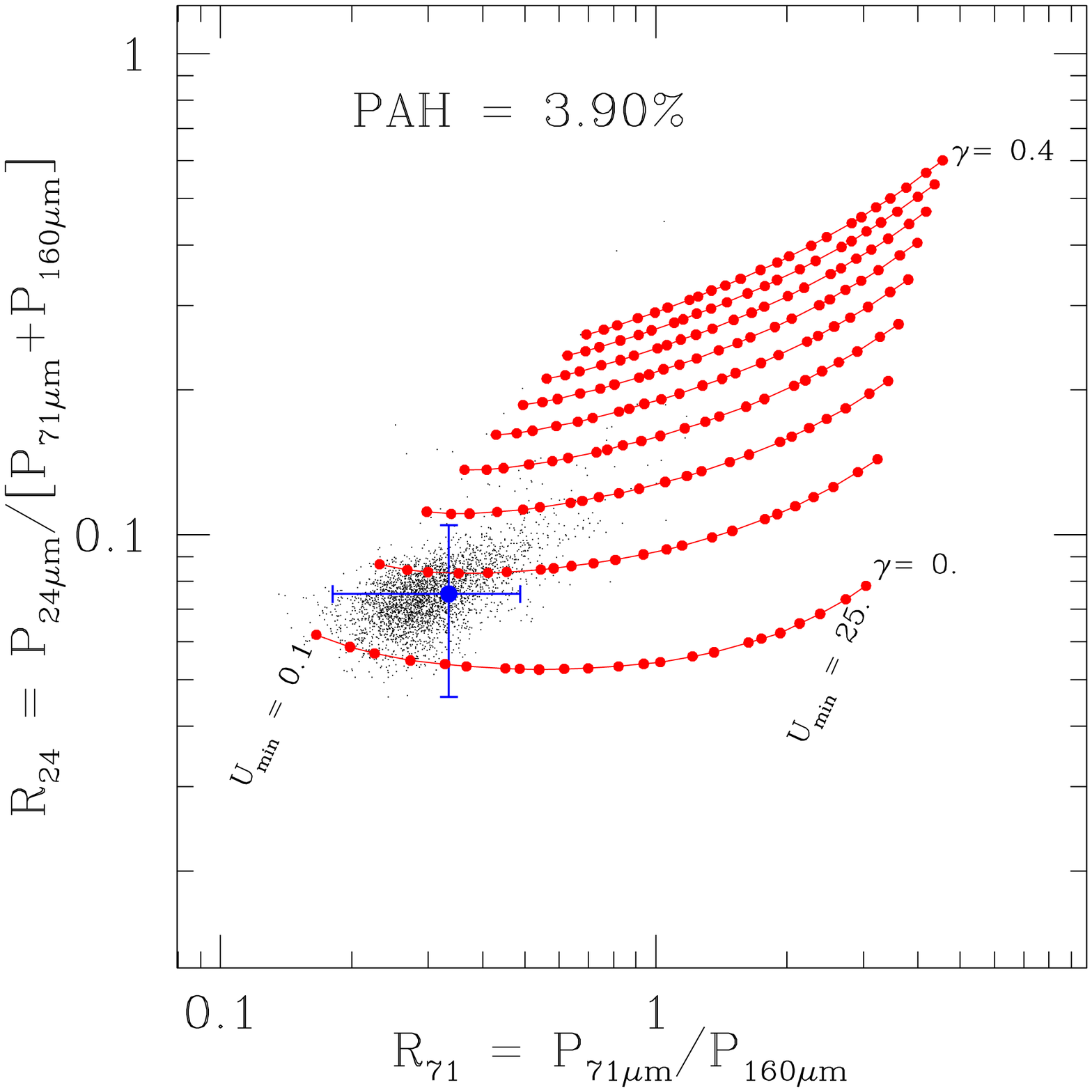}
  \includegraphics[width=9.cm]{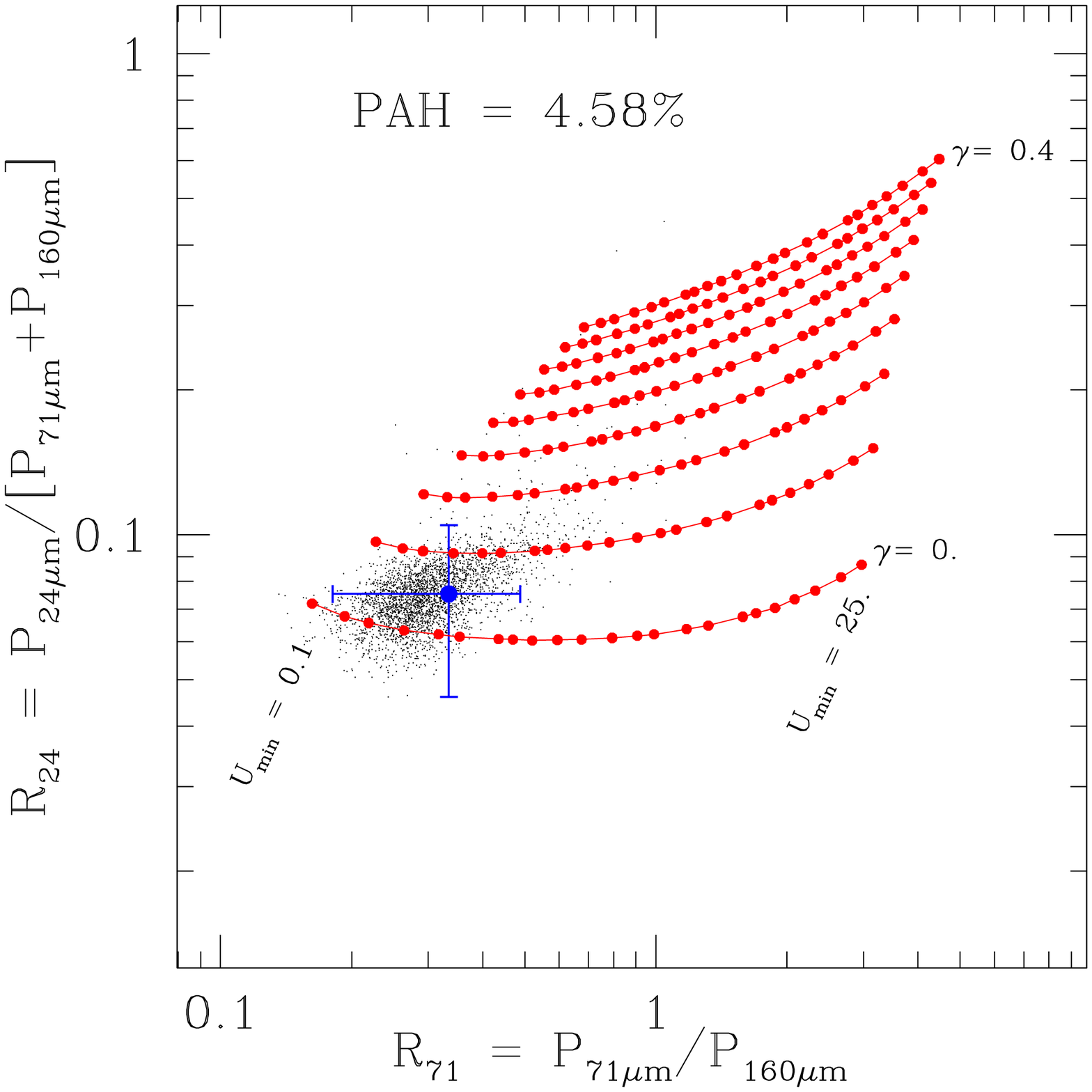}
  \caption{Ratio $R_{24}$ of the 24$\U{\mu m}$ flux normalized to FIR
    emission against $71\U{\mu m}$ to 160$\U{\mu m}$ flux ratio
    $R_{71}$. The blue point is the ratio correspondent to the total
    integrated flux measurement of \object{M31}. Each horizontal red
    line shows the ratio values predicted by the models of Draine \&
    Li~(2007) for different values of the minimum radiation field
    intensity $U_\M{min}$ and of the $\gamma$ parameter (see
    text). The three panels show the results for models with different
    abundances of PAHs as indicated by the labels.}
  \label{fig:r24}
\end{figure*}

In Fig.~\ref{fig:r24} we show the ratio $R_{24}$ of the $24\U{\mu m}$ flux
(normalized to the total infrared emission) against the ratio $R_{71}$,
a similar diagram of Fig.\ref{fig:r8}, but for the
24$\U{\mu m}$ map, which is better diagnostic to the radiation field
intensity (Draine et al.~2007). We considered in this case only models 
with PAHs abundances $>3.19\%$ since for the analyzed regions  
these are the models that better agree with observations 
(Fig.~\ref{fig:r8}). For the bulk of the points, and for the total
integrated flux measurement $\gamma$ can be in general considered
$\leq 0.05$. Using the equation:

\begin{equation}
f_{PDR}=1.05(R_{24}-0.14R_{8}-0.035)^{0.75}
\end{equation}

\noindent
taken from Draine \& Li~(2007), and the observed ratios $R_{8}$ and
$R_{24}$ defined above, we determined the fraction of the total dust
luminosity that is radiated by dust grains in regions with $U>10^2$ 
($f_{PDR}$) to be lower than $\sim 4\%$.

\section{Which sources are responsible for the dust heating in \object{M31}?}
\label{sec:pop}

\begin{figure*}
  \centering
  \includegraphics[width=15cm]{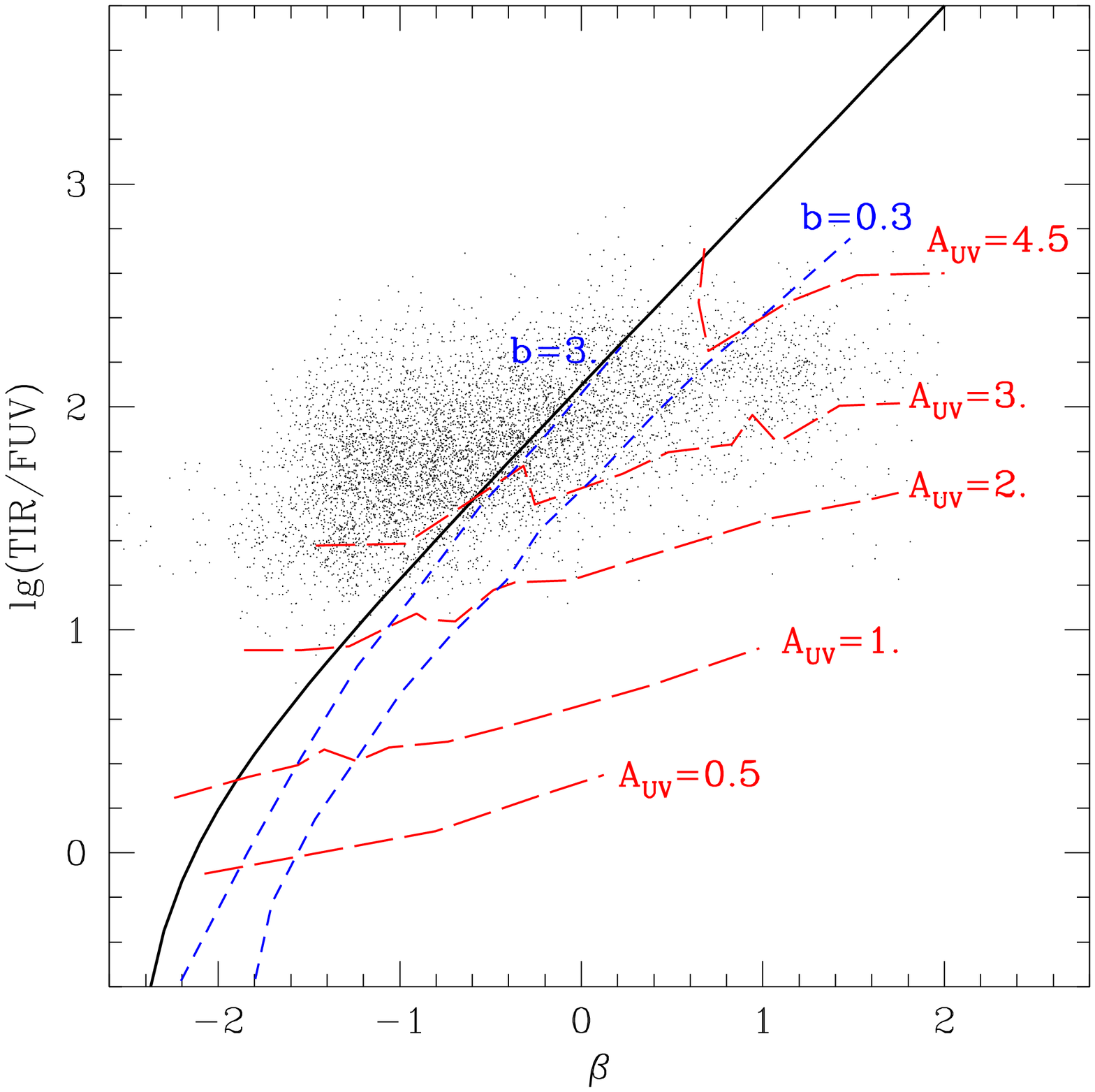}
  \caption{ $(\TIR/\FUV)$ logarithmic ratio against $\beta$, the
    UV-slope, for each region with $\rm S/N>1$ analyzed in the $40\arcsec$/px
    resolution maps. The thick continuous line denotes the best fit of Kong
    et al.~(2004) for their sample of starburst galaxies. The long 
    dashed red lines and the short dashed blue lines are constant
    attenuation and constant ratio of present to past-averaged star formation
    rate lines as given by Kong et al.~(2004).}
  \label{fig:tir_fuv_nuv}
\end{figure*}

The analysis of the infrared spectrum presented in
Sec.~\ref{sec:models} allows to conclude that the infrared spectrum of
\object{M31} can be explained by diffuse dust emission. Nevertheless,
on the basis of that, we cannot understand for which stellar populations
are responsible for the dust heating process. In
particular we want to know where the dust in \object{M31} is
predominantly heated by young populations ($\U{Age}<1\U{Gyr}$) 
or where older populations ($\U{Age}>1\U{Gyr}$) play an important role 
in the dust-heating process as well.  
As shown below the answer to this
question is of crucial impact for the deduction of several important
parameters, e.g. attenuation and SFR, and thus has to be addressed as
a central issue.

In the following we used the observational maps of \object{M31}
resampled at 40$\arcsec/\M{px}$ as described in Sec.~\ref{sec:data},
and considered all the regions in which the $S/N>1$ in all the maps.

Also it should be noted that the results we obtained are based on the 
theoretical models of Kong et al.~(2004) and Cortese et al.~(2008). It is 
worth saying that alternative interpretations could exist of the observational
data we present. In particular models considering
modifications of the canonical attenuation laws have been proposed (e.g. Inoue 2005, 2006
see later in this Section). 

The model of Kong et al.~(2004) is based on the Bruzual \& Charlot~(2003) population 
synthesis code, it assumes an exponential star formation history (SFH) and a power-law
absorption law, distinguishing between young ($<10^{7}\U{yr}$) stars
embedded in their birth clouds and older stars migrated into the ISM. The model of Cortese et al.~(2008) 
is also based on the Bruzual \& Charlot~(2003) population syntesis code, but is assumes
an LMC attenuation law, and that both stars and dust were homogeneously 
distributed in a plane parallel (sandwich) geometry. They also adopted a Star Formation History
'a la Sandage' in the formalism of Gavazzi et al. (2002):  

\begin{equation}
SFR(t,\tau)=\frac{t}{\tau^2}exp\left(-\frac{t^2}{2\tau^2}\right)
\end{equation}

\noindent
where SFR is the Star Formation Rate per unit mass, t the age (in Gyr) of the galaxy
(assuming t=$13\U{Gyr}$ at present epoch), and $\tau$ is the time (in Gyr) at which 
the star formation rate reaches the highest value over the whole galaxy history 
(note that in this notation the present age of the stellar populations born at time $\U{t}=\tau$
is given by $\U{age}=13 \U{Gyr}-\tau $).
As stated by the same authors, $\tau$
should be considered as a proxy for the shape of the SED rather than being used to 
derive an exact indication of the age of the underlying stellar populations. We calculated
for each value of $\tau$, the time $\U{t^{\star}}$ where the 
SFR reaches half of its maximum value (for $\U{t}>\tau$) over the whole galaxy history. 
Also the age of the stellar populations correspondent to $\U{t}=\U{t^{\star}}$ must not be 
considered in a strict manner but it can be considered as an indication of the age of the youngest
stellar populations that are significantly contributing to the SED, given that
the contribute of next generations of stars becomes increasingly less important in order to explain the observed SED. 
For example, for $\tau=8\U{Gyr}$ we obtain $\U{t^{\star}\eqsim13.5\U{Gyr}}$.
Assuming for the present epoch t=13$\U{Gyr}$ as in Cortese et al.~(2008),
we can expect that young populations ($\U{age}<1\U{Gyr}$) contributed 
significantly to the observed SED. For $\tau=5\U{Gyr}$ we obtain instead 
$\U{t^{\star}\eqsim9.6\U{Gyr}}$ which implies that stellar populations 
contributing to the SED should have an $\U{age}>3\U{Gyr}$.

It should be also remind that the predictions of these models should not be considered
reliable in regions where old stars are likely to dominate the UV flux
(mostly in the bulge region for \object{M31}).

At first we investigated the $\TIR/\FUV$ vs. $\beta$ (the UV slope) relationship
for the analyzed regions as shown in Fig.~\ref{fig:tir_fuv_nuv}. The
solid line represents the fit obtained by Kong et al.~(2004) for the
$50$ starburst galaxies in their sample, which represents the so
called $IRX-\beta$ relationship for starburst galaxies (Meurer et
al.~1999). We adopted the same definition of Kong et al.~(2004) to
calculate $\beta$:

\begin{equation}
\beta\,=\,\frac{lg(\overline{f}_{FUV})-lg(\overline{f}_{NUV})}{lg(\lambda_{FUV})-lg(\lambda_{NUV})}
\end{equation}

\noindent
where $\lambda_{FUV}\,=\,1516\,\AA$ and $\lambda_{NUV}\,=\,2267\,\AA$ 
are the effective wavelengths of the far-ultraviolet and near-ultraviolet filters 
on board of GALEX and $\overline{f}_{FUV},\overline{f}_{NUV}$ are the 
mean flux densities (per unit wavelength) through these filters.

The small black points are the
values of the $\TIR/\FUV$ and $\beta$ obtained in the analyzed
regions of \object{M31}. A positive correlation is visible, but it
does not follow the relation found in starburst galaxies. As can
be expected naively, going towards negative slopes and bluer colors
our observations imply similar $\TIR/\FUV$ and $\beta$ values with respect
to the starburst galaxies relationship whereas redder regions behave
differently showing typically lower $\TIR/\FUV$ for equal
$\beta$. This is in line with Kong et al.~(2004), who showed the
results coming from the analysis of their sample of normal
star-forming galaxies as well and obtained that these objects
typically have lower values of $\TIR/\FUV$ for fixed UV-color with
respect to what predicted by the $(IRX-\beta)$ relation, with a large
scatter of values.  Other authors have recently questioned the
application of the $(IRX-\beta)$ relationship for normal star-forming
galaxies, suggesting that such systems suffer from lower dust
attenuations with respect to what could be inferred from the
$(IRX-\beta)$ relation (Salim et al. 2007).  

Following the notation of Kong et al.~(2004) we plot in Fig.~\ref{fig:tir_fuv_nuv} 
the lines of constant ratio of present to past-averaged star formation rate assuming an
an exponentially declining star formation history ($b$ parameter, blue short-dashed lines)
and of constant attenuation ($\rm A_{UV}$, red long-dashed lines). 

This comparison suggests that in the \object{M31} regions population gradients are present but
there are similar amounts of dust attenuation, being our observational points
elongated along the direction indicated by the bands of equal
attenuation.

As color variations seem due to a spread of ages, it is more convenient to 
adopt a larger color baseline like the $(\FUV-i_\M{SDSS})$\footnote{
We have decided to use here the $(\FUV-i_\M{SDSS})$ color as the $2MASS$
images of \object{M31} obtained during the $2MASS\,Extended\,Survey$
did not appear sufficiently deep for our purposes.} color. Cortese
et al.~(2008) provided a detailed discussion of the age dependence of
the $\TIR/\FUV$ ratio, colors and attenuations.  In accordance with
their results, for $(\FUV - i_\M{SDSS}) \sim 4.3$ the energy absorbed by
the dust at $\lambda < 4000\,\AA$ is approximately equal to the energy
absorbed at $\lambda > 4000\,\AA$ (as inferred from their Fig.~2 and
Table~1). This corresponds to $\U{t^{\star}\simeq}8.5\U{Gyr}$ as defined above,
thus the stellar populations that are contributing to the SED have an age$>4.5\U{Gyr}$.

\begin{figure*}
  \centering
  \includegraphics[width=15cm]{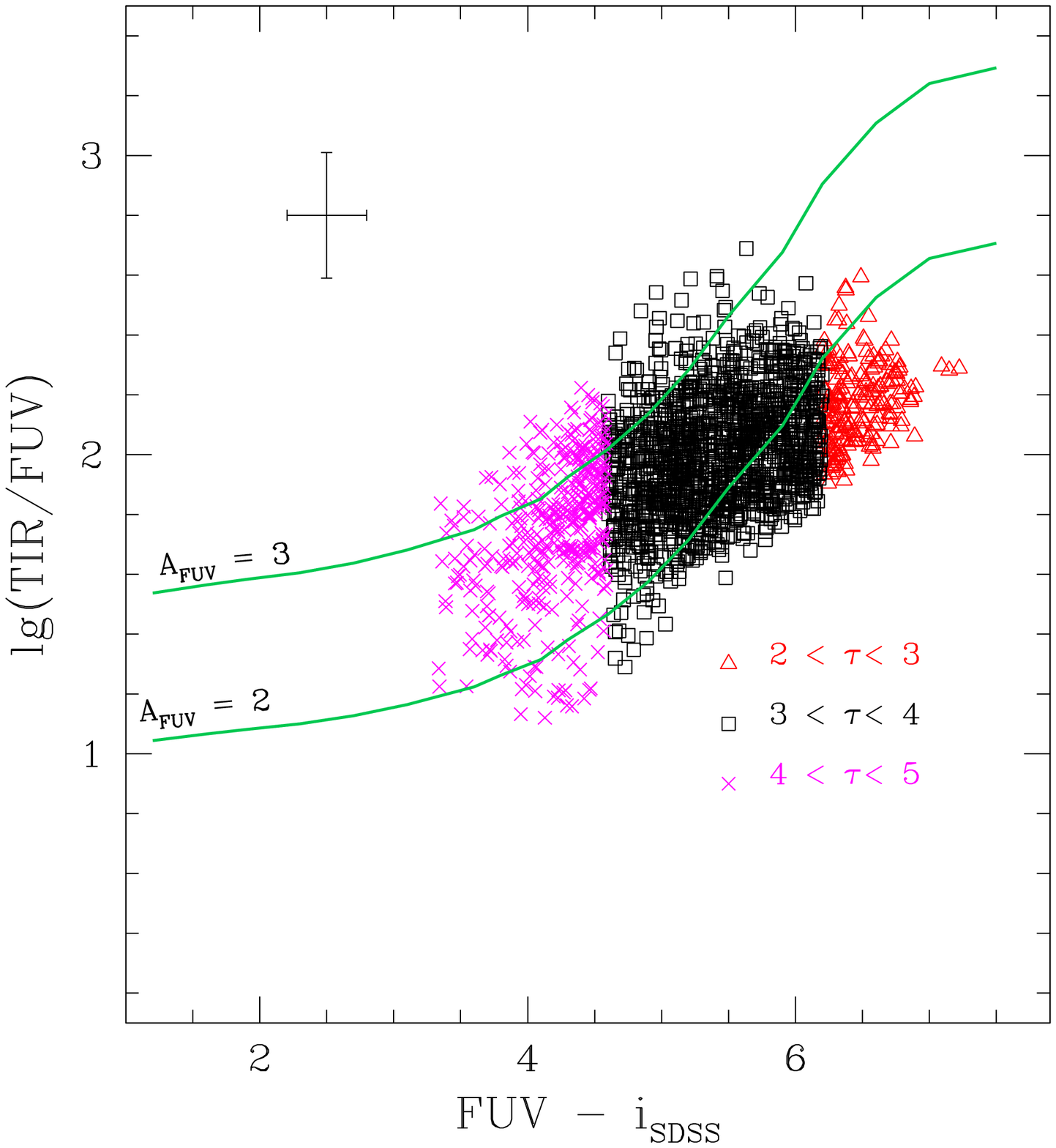}
  \includegraphics[width=8cm]{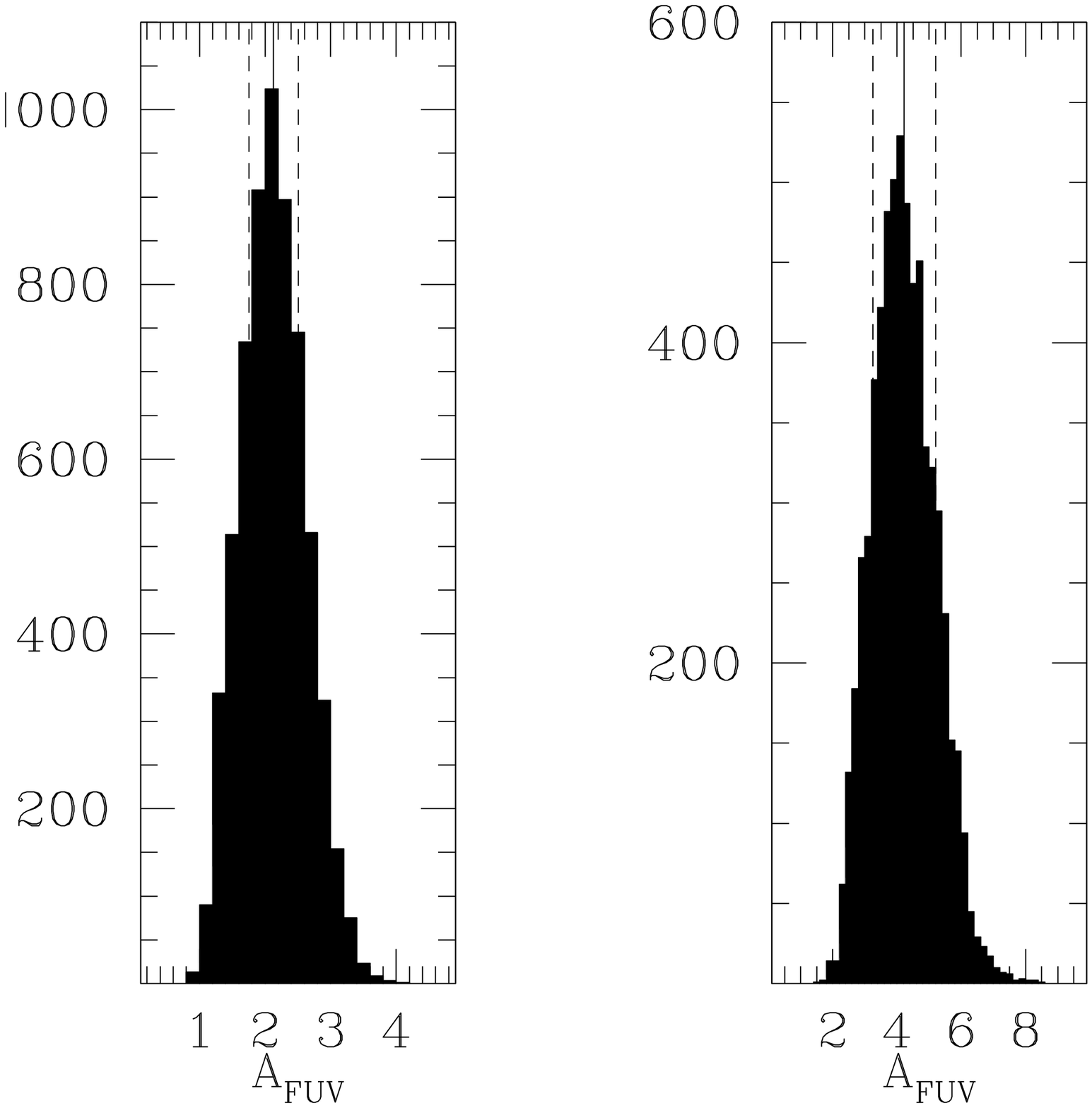}
  \caption{Upper panel: $(\TIR/\FUV)$ logarithmic ratio against
    $(\FUV-i_\M{SDSS})$ color for regions with $S/N\,>\,1$ across the
    10 kpc ring ($8kpc<r<12kpc$). The colors denote the different
    values of $\tau$ 
     (the time expressed in Gyr at which the star formation rate reaches 
     the highest value over the whole galaxy history, assuming $13\U{Gyr}$ 
     at present epoch) obtained for each region
    following the recipes of Cortese et al.~(2008). The two solid
    lines are two lines of constant attenuation.  The errorbar on the
    top-left corner shows the mean observational errors.  Lower
    panels: histograms of ultraviolet attenuations derived following
    the recipes of Cortese et al.~(2008, left panel) and assuming the
    starburst scenario (right panel). Mean values and $1\,\sigma$
    uncertainties are denoted by solid and dashed vertical lines and
    are $A_\FUV\,=\,2.1\,\pm\,0.4$ mag and $A_\FUV\,=\,4.2\,\pm\,1.0$
    mag for left and right panel respectively.  }
  \label{fig:tir_fuv}
\end{figure*}

In Fig.~\ref{fig:tir_fuv} we show the $\TIR/\FUV$ ratio against the
$(\FUV-i_\M{SDSS})$ color, for regions with $S/N>1$ across the 10kpc ring
($\rm 8kpc<r<12kpc$). The positive correlation between these variables is
still clearly visible also in this diagram. Moreover the color range
$3.5 < (\FUV-i_\M{SDSS})< 7$ demonstrates that in a large number of
regions old populations dominate the dust heating process.  As shown
in Fig.~\ref{fig:tir_fuv} (lower left panel), the mean $A_\FUV$ in
the analyzed regions of \object{M31} derived with this method is
$A_\FUV = 2.1 \pm 0.4\U{mag}$ where the error is the standard
deviation.  If the starburst scenario is considered, we obtain the
result shown in the lower right panel of Fig.~\ref{fig:tir_fuv}, where
the mean attenuation is $A_\FUV = 4.2 \pm 1.0\U{mag}$. Neglecting
the dust heating age dependence would imply a mean attenuation $100\%$
larger and a scatter $150\%$ larger, since in this case the large
color changes would totally be attributed to the reddening.

In Fig.~\ref{fig:tir_fuv}, we overplotted two lines of constant
attenuation correspondent to $A_\FUV\,=\,2$ and $A_\FUV\,=\,3$ as
indicated by the labels. The shape of these lines reflects the age
dependence of the dust heating as modeled by Cortese et
al.~(2008). Actually for blue $(\FUV-i_\M{SDSS})<4$) or red colors
$(\FUV-i_\M{SDSS})>7$ the attenuation lines tend to be parallel to
the color axis, and thus independent from the color. This is due to
the fact that in these color ranges the intrinsic spectral energy
distribution of the populations that are heating the dust is not
expected to change appreciably with age to the very young (blue) or
very old (red) side. On the other hand in the range comprised between
$4<(\FUV-i_\M{SDSS})<7$, which is actually where most of our
observations fall, age variations contribute significantly to color
changes, and older populations appear intrinsically redder thus
shifting the $\TIR/\FUV$ ratio towards larger values for equal amounts
of attenuation with respect to bluer regions. In the following we
tried to disentangle the effects of the age and of the reddening on
the observed colors and $\TIR/\FUV$ ratios.

At first we converted the colors to the values of $\tau$ 
(the time expressed in Gyr at which the star formation rate reaches the highest value over the
whole galaxy history, assuming $13\U{Gyr}$ at present epoch) 
by means of:

\begin{equation} 
  lg(\tau) = -0.073\,(\FUV\,-\,i_\M{SDSS}) + 0.96
\end{equation}

taken from Table~2 of Cortese et al.~(2008). The colors in
Fig.~\ref{fig:tir_fuv} are codified in function of the value of
$\tau$. We obtained that $83\%$ of the analyzed regions across the
spiral-ring have $(\FUV - i_\M{SDSS}) > 4.3$ and thus values of $\tau <
4.4\U{Gyr}$.  Hence according to the results of Cortese et al.~(2008) in
$83\%$ of the analyzed regions the dust absorbs more than $50\%$ of
the energy at $\lambda > 4000\,\AA$. As stated above this 
implies that in these regions stellar populations responsible for
the dust heating should be at least a few Gyr old. 
This could provide a good agreement to
the low mean intensity of the radiation field we found from the
analysis of the infrared spectrum of \object{M31}, although a small value
of $U$ could be obtained also by young stars with a low SFR.
It is important to remind that these results do not imply that younger populations are not present
at all, but rather that their contribute is not dominant in these regions.
If, on the contrary one tries to explain the observed colors assuming
that only young populations (age$<1\U{Gyr}$) are heating the dust, one should also 
admit much larger attenuations for normal galaxies, as demostrated above and
for the same distribution of star and dust as in Cortese et al.~(2008).
Otherwise it is clear that in order to confirm these results more accurate 
models are needed. Some other considerations regarding the limitation of the Cortese et al.~(2008) 
approach are presented shortly below and in the next Section.

\begin{figure*}
  \centering
  \includegraphics[width=15cm]{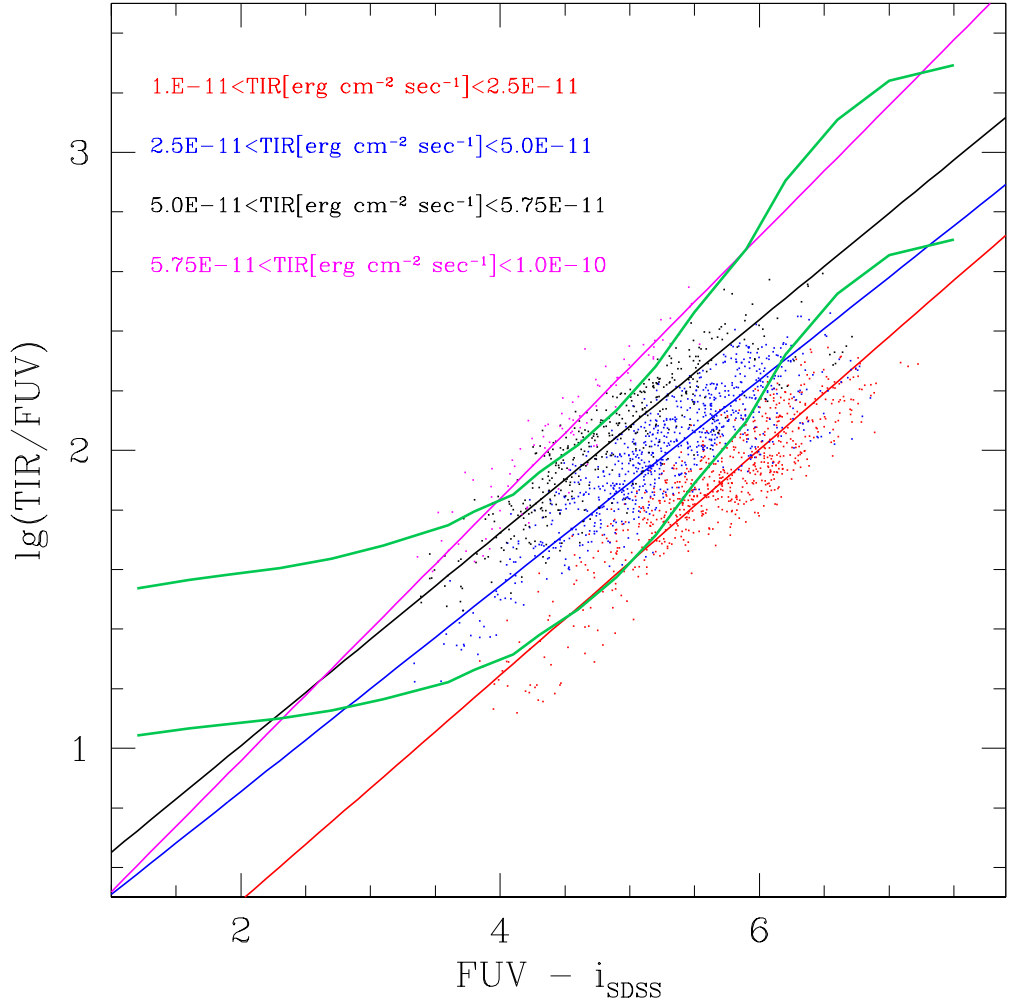}
  \caption{$\TIR/\FUV$ logarithmic ratios vs. $(\FUV-i_\M{SDSS})$
    colors for regions with $S/N>1$ across the 10kpc ring
    ($8kpc<r<12kpc$). Different colors and labels denote regions of
    constant TIR emission. Red, blue, black and magenta lines indicate
    the best fitting linear regression models of the data in each TIR
    emission interval.  Green curves are two constant attenuation
    lines accordingly to the recipes of Cortese et al.~(2008) as shown
    in Fig.~\ref{fig:tir_fuv}.}
  \label{fig:tir_grow}
\end{figure*}

  From another point of view we show in Fig.~\ref{fig:tir_grow} the
  same diagram of Fig.~\ref{fig:tir_fuv} separating regions of low and high TIR emission as
  indicated by the labels and the colors. We considered 4 equal bins of
  TIR emission comprised between $10^{-11}-10^{-10}\U{erg}\U{cm^{-2}}\U{sec^{-1}}$. From
  Fig.~\ref{fig:tir_grow} it appears that moving towards redder colors
  along the regions of constant TIR emission, the $\TIR/\FUV$ ratio
  increases. As the TIR emission is constant, the FUV emission must
  decrease. Nevertheless, if the lower FUV emission was due to larger
  attenuations, the TIR emission should have increased accordingly,
  which is not observed. Thus, the color change must be due to an
  intrinsic lower emission in the FUV band, that could be
  indicative of the presence of a population gradient. 
  Strictly speaking this interpretation is valid for 
  an homogeneous layer of stars embedded in an homogeneous optically thin layer
  of dust. In a real galaxy dust and stars (especially recently formed stars)  
  are likely to have more clumpy and inhomogeneous distributions. Otherwise, 
  studying two of the largest dust clouds listed in the 
  \object{M31} Atlas (Hodge 1980), Hodge \& Kennicutt (1982)
  derived average extinctions of $\U{A_B}=0.35$ mag (with a maximum value
  of $\U{A_B}=0.60$ mag) and $\U{A_B}=0.43$ mag (maximum  $\U{A_B}=0.76$ mag) for D307
  and D441 respectively, indicating that the majority of all the other clouds 
  should have smaller amounts of extinctions. Other studies confirm the idea 
  that the interstellar medium in \object{M31} is
  generally optically thin. Fan et al.~(2007) derived reddening values towards
  443 \object{M31}'s catalogued globular clusters obtaining that more than
  half of them are affected by a reddening $\U{E(B-V)}<\sim0.2$ mag with an average
  value of $\U{E(B-V)}=0.28^{+0.23}_{-0.14}$ (see also Barmby et al.~2000). 
  Regarding the homogeneity of the dust distribution, Inoue~(2005, 2006) investigating the attenuation
  law resulting from clumpy spatial distributions of dust and (young) stars,
  proposed different explanations for the redder $UV$ colors of normal galaxies 
  with respect to starburst galaxies for fixed FIR to FUV ratios, which are not based 
  on population gradients. These alternative models can be summarized as: (i) models 
  where the attenuation law has a steeper dependence on wavelength than the canonical
  attenuation laws as a result of an age-selective attenuation of young stars in clumpy
  structures with respect to old stars; (ii) models where the attenuation law has no
  attenuation bump at $2175\AA$; (iii) models with a bump at $2175\AA$, but with 
  smaller albedo for shorter wavelengths (except for the bump range).
  While we did not try to apply these models to our data, we observe that at least the
  first two scenarious do not seem to fit the case of \object{M31}.  
  Bianchi et al.~(1996) derived that the \object{M31} extinction curve is very
  similar to the average Galactic extinction law and a possible reduction of the $2175\AA$ bump is
  significant only at the 1$\sigma$ level. Similar results were reached also in previous studies 
  (e.g. Walterbos \& Kennicutt 1988). As for the third class of models further studies 
  of the wavelength dependence of the albedo are necessary. We also observe that these considerations
  do not imply that clumpy structures with young stars embedded do not exist in \object{M31}, 
  but it could be that most of the dust in \object{M31} is not located in these structures. From counts of dark nebulae 
  across the disk of \object{M31} Hodge \& Kennicutt (1982) obtained that the major dust
  lanes visible in the optical should account for $\sim15\%$ of the total dust mass content. 
  Recent results by Nieten et al.~(2007) show a good correlation between the 
  most prominent dust lanes and dense molecular clouds traced by strong CO emission lines but
  concluded that molecular gas in \object{M31} is only $\sim7\%$ of the total
  neutral gas content in \object{M31}, and that dust appears correlated also with
  atomic gas in \object{M31} which is distributed in more 
  extended regions out of the densest clouds in the spiral-ring structure.
  Moreover in Sec.~\ref{subsec:massestimate} we derived that only a fraction $<7\%$ of the TIR comes
  from the $hot$ dust component emission at 24$\mu$m associated with young star forming regions.
  Finally, it is unlikely that a large fraction of the \object{M31}'s dust mass is in very cold dark clouds
  ($\U{T}\,<\,16\,\U{K}$), as the total \object{M31} dust mass we derived in Sec.~\ref{subsec:massestimate} 
  from dust diffuse emission models was close (1$\%$-5$\%$) to the estimate based on neutral gas measurements. 
  We conclude that the interpretation of Fig.~\ref{fig:tir_grow} in terms of 
  population gradients must be considered carefully and in need of further investigation, 
  but it appears reasonable for regions dominated by diffuse dust emission away from
  large star-forming complexes (in this regard remember the criteria we applied
  in Sec.~\ref{sec:data} against the selection of HII regions).  
  In fact the attenuation along bands of
  equal TIR emission appears to decrease towards redder colors
  accordingly to the predictions of Cortese et al.~(2008).  This 
  could be also in agreement with the idea of a population gradient, as
  older populations would tend to be intrinsically less attenuated
  than younger ones, because of their intrinsically redder spectral
  energy distribution, at least in the region where the age effect is
  maximal $4<(\FUV - i_\M{SDSS})<7$. We note nevertheless that the
  observed color could be also biased towards redder values especially
  in the bulge region where the geometry of the dust and star is
  clearly deviating from the simplyfied sandwich model of Cortese et
  al.~(2008). We will discuss such potential bias effect in more
  details in the next Section. Another intresting consideration
  regarding the diagram of Fig.~\ref{fig:tir_grow}, is that moving
  vertically towards larger $\TIR/\FUV$ ratios the TIR emission
  increases (whereas the $\FUV$ emission remains almost constant), which
  should imply larger attenuation values because of the large absorbed
  energy for a fixed observed $\FUV$ energy, which is what the model
  predicts.  We have thus the following interpretation of the diagram
  in Fig.~\ref{fig:tir_grow}.  While the large color variation of the
  analyzed regions seems to be due to the gradient of populations that
  are heating the dust, the spread of the $\TIR/\FUV$ values for a fixed
  color seems instead due to the differential reddening of regions
  with similar underlying stellar populations.

\begin{figure*}
  \centering
  \includegraphics[width=15cm]{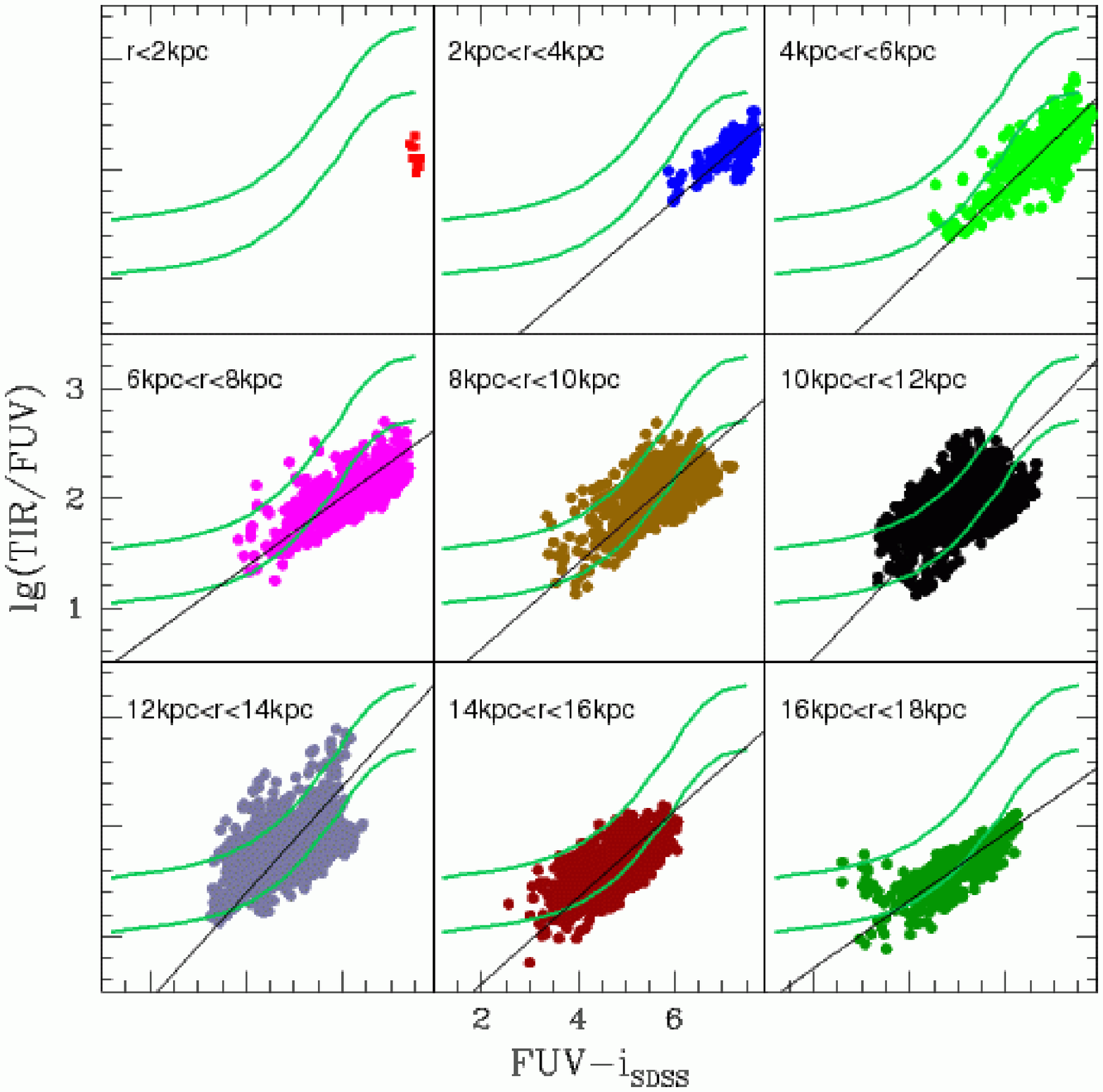}
  \caption{Logarithmic $\TIR/\FUV$ ratios
    against $\FUV-i_\M{SDSS}$ colors for regions inside each
    annulus. Thin black solid and dashed lines are best fitting linear
    models of the data as explained in the text.  Green curves are
    constant attenuation lines accordingly to the recipes of Cortese
    et al.~(2008) as shown in Fig.~\ref{fig:tir_fuv}.}
  \label{fig:valuemap}
\end{figure*}

  We further investigated the spatial
  distribution of the $A_\FUV$ across the disk of \object{M31}. We
  considered nine $2\U{kpc}$-wide radial bins from the center of
  \object{M31} up to $18\U{kpc}$ as shown in Fig.~\ref{fig:sn}, 
  and imposed a $S/N > 1$.  In Fig.~\ref{fig:valuemap} we present the logarithmic $\TIR/\FUV$
  ratios against the $(\FUV-i_\M{SDSS})$ colors obtained in each
  radial bin. In general it appears that these observable quantities
  are positively correlated across all the disk of \object{M31}. Only
  in the innermost annulus it was not possible to verify the positive
  correlation of these variables because of the few regions
  analyzed. We fit the $\lg(\TIR/\FUV)$ ratios against to
  $(\FUV-i_\M{SDSS})$ colors in each radial bin with a linear model:
  $\lg(\TIR/\FUV) = a\,(\FUV-i_\M{SDSS}) + b$ . We considered
  separately the uncertainties along the $x$ and $y$ axis and took the
  mean value of the derived parameters and their semi-difference as
  estimates of the best model parameters ($a$ and $b$) and
  uncertainties. In Tab.~\ref{tab:radial_plots} we reported the result
  of the fit along with other useful quantities.

\begin{figure*}
  \centering
  \includegraphics[width=15cm]{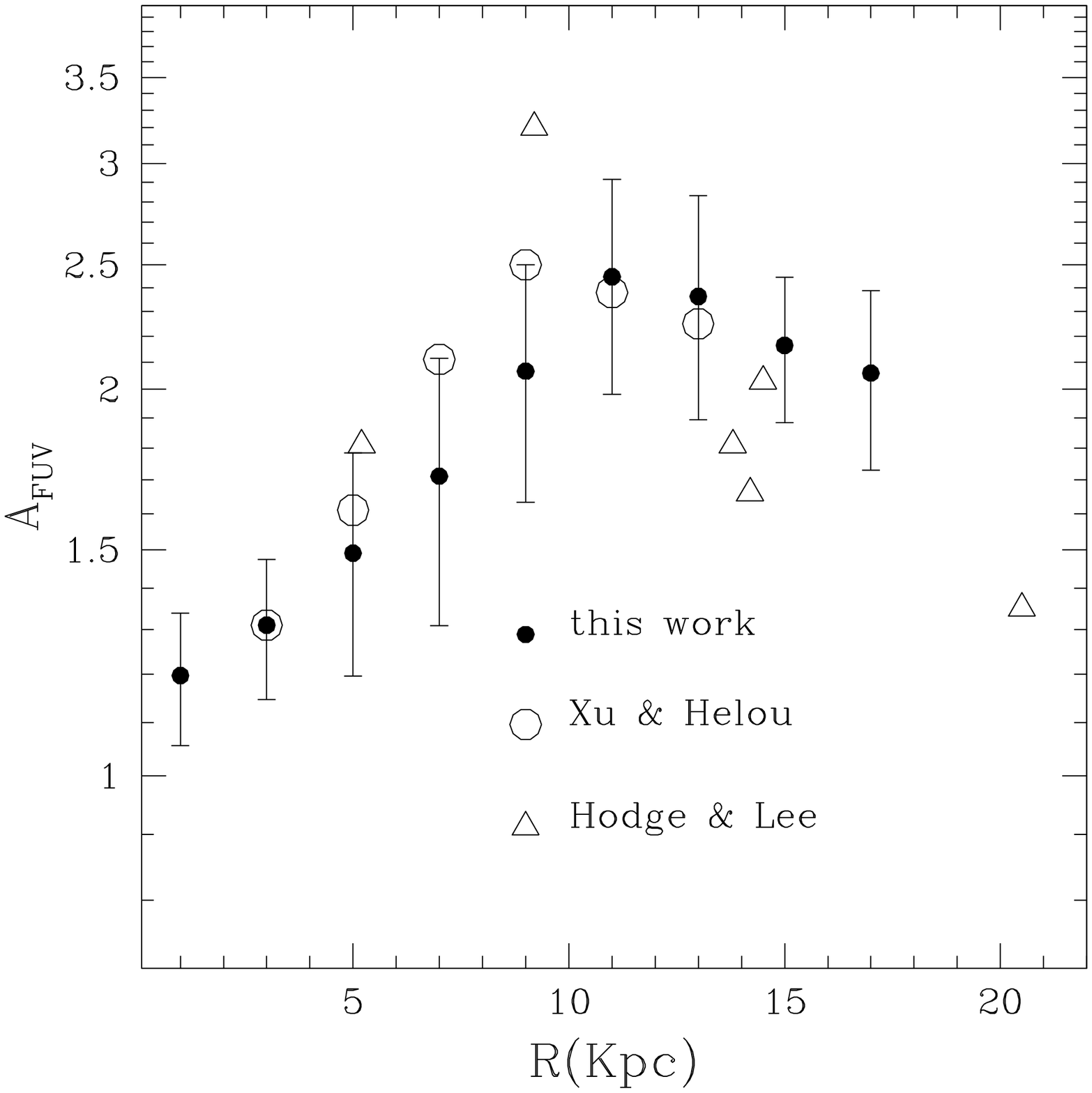}
  \caption{Average FUV ($\U{\lambda_{eff}}=1516\AA$) attenuations in 2kpc radial bins from the center of
    \object{M31} to $\sim$21kpc obtained by us (filled circles), Xu \&
    Helou (1996, open circles). Open triangles represent the result of
    Hodge \& Lee (1998) obtained analyzing 5 fields at different
    galctocentric distances.}
  \label{fig:rad_profiles}
\end{figure*}

  As shown also in Fig.~\ref{fig:rad_profiles} the mean attenuation
  in each radial bin varies reaching a maximum near the 10kpc ring,
  while it seems to decrease faster towards the inner regions of
  \object{M31} than in the outer regions.

\begin{figure*}
  \centering
  \includegraphics[width=18cm]{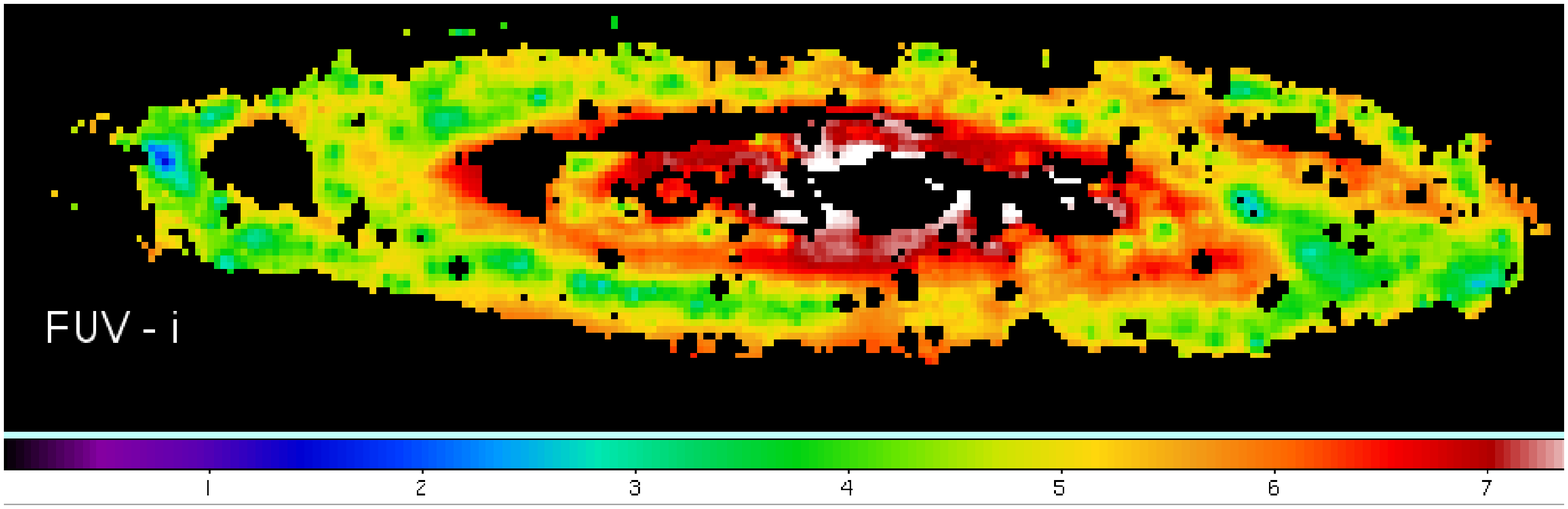}
  \includegraphics[width=18cm]{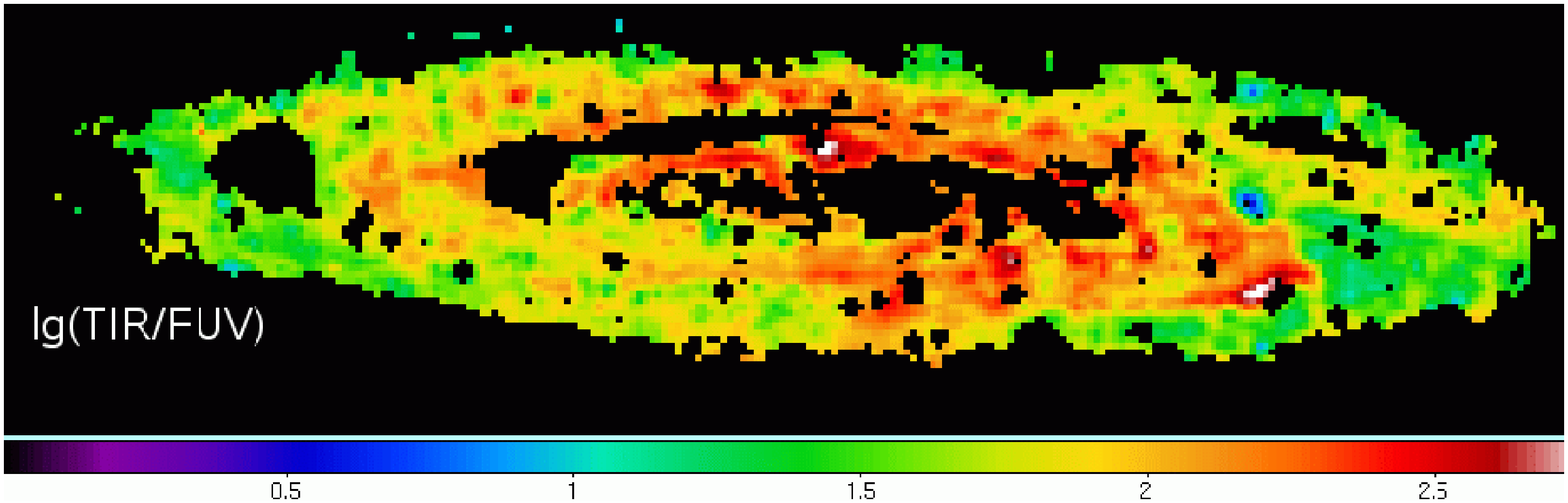}
  \includegraphics[width=18cm]{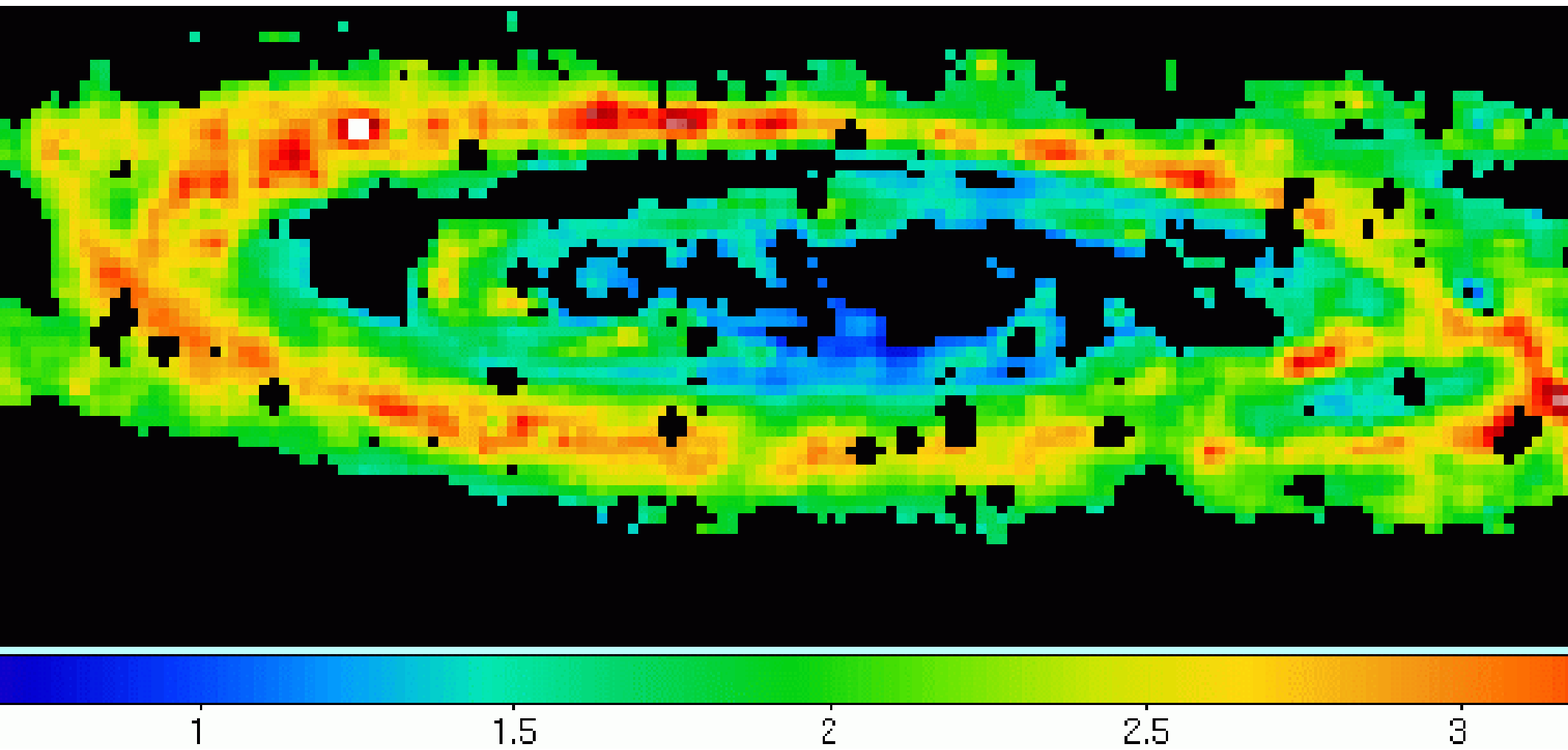}
  \caption{From top to bottom: $(\FUV-i_\M{SDSS})$ map, logarithmic
    $\TIR/\FUV$ map, and $A_\FUV$ attenuation map. Color bars indicate
    the range of values spanned by the quantities using a linear
    scale. The first two maps are derived combining the $\FUV$,
    $i_\M{SDSS}$ and all the infrared maps used in this work. The
    regions shown here have $S/N\,>\,1$, $\FUV/NUV\,>\,0.3$ and
    $1\,<\,F^\M{ns}_{8\mu m}/ F^\M{ns}_{24\mu m}\,<\,10$, and avoid bright
    sources detected in the $i_\M{SDSS}$ band (black dots). The last
    map shows the attenuations $A_\FUV$ derived using the recipes of
    Cortese et al.~(2008) and the two above maps. The attenuation is
    maximum in the 10kpc ring where it reaches a mean value of
    $2.5\,\pm\,0.7$ mag, and declines more strongly towards the inner
    regions of the galaxy than towards the outer regions.}
  \label{fig:colmaps}
\end{figure*}

  These results are also shown as the color maps of
  Fig.~\ref{fig:colmaps}. While as dicussed above the model 
  adopted here could be less reliable in the inner part of the galaxy due
  to deviations of the dust and star distribution from the assumed
  geometry, the trend and the values of the mean attenuations derived
  with this method are in agreement with previous studies as discussed
  in the next Section.  

The existence of a correlation between the $\TIR/\FUV$ ratios and the
colors that we have found needs a final remark. Actually it is
important to keep in mind that the theoretical calculations of Kong et
al.~(2004) and Cortese et al.~(2008) do not necessarily imply the
existence of any correlation between these observables. Nonetheless we
should remind that in this case we are studying regions inside $one$
galaxy, and thus we can expect a more uniform and homogeneous
behaviour.

\section{Discussion}
\label{sec:discussion}

  The models of Cortese et al.~(2008) assume a simple plane
  parallel (sandwich) geometry for the dust and star
  distribution. Even if the fractional scale height between the dust
  and the stars is allowed to vary with wavelength no radial
  dependence is assumed. The method solely grounds on the relationship
  between two observed emission ratios and can by this be expected to
  be highly prone to light contamination from distant sources and a
  different geometry. Since, if the geometry is not plane parallel the
  radiation field experienced by the dust can be significantly
  different from what is naively measured by the outside observer, the
  observed color from which the age of the underlying population is
  estimated could be biased towards redder values in regions where
  older stellar populations are less embedded in the dust. This bias
  should be strongest in the inner regions of the galaxy as the result
  of the bulge light contamination. Regions with $r<8\U{kpc}$ in
  Fig.~\ref{fig:valuemap} reach redder colors and higher TIR/FUV ratios
  with respect to the outer regions. 
  For $\rm r>12kpc$ the color range remains almost constant although the   
  TIR/FUV ratio decreases.
  In presence of the above mentioned observational bias the
  value of the attenuation derived from the model could have been
  underestimated in the inner regions. Moreover in the innermost radial
  bin ($r<2\U{kpc}$) old stars can contribute a significant fraction
  of the GALEX-UV emission, and thus the model
  prediction should not be considered reliable in that region.

\begin{figure*}
  \centering
  \includegraphics[width=8cm]{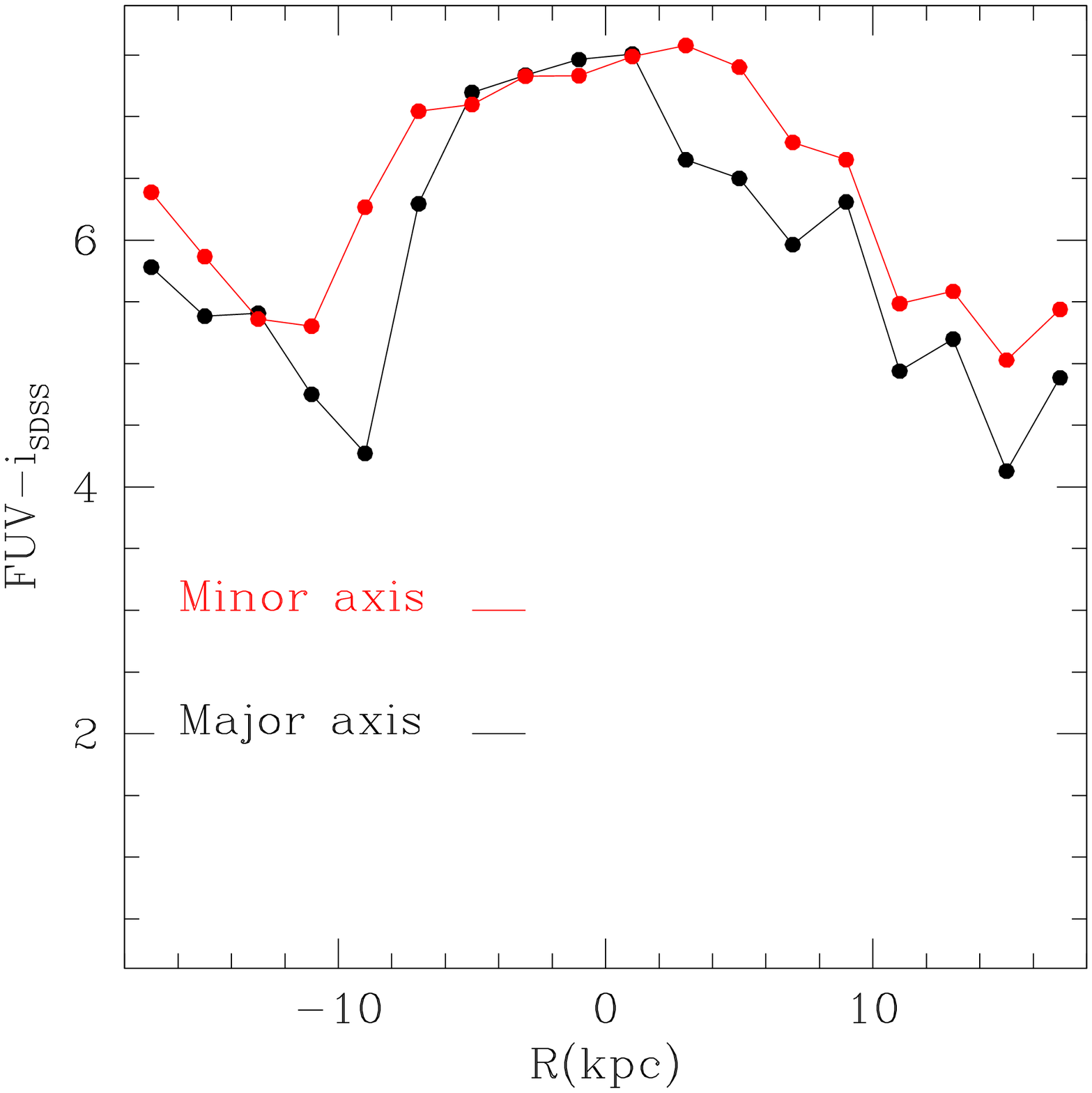}
  \includegraphics[width=8cm]{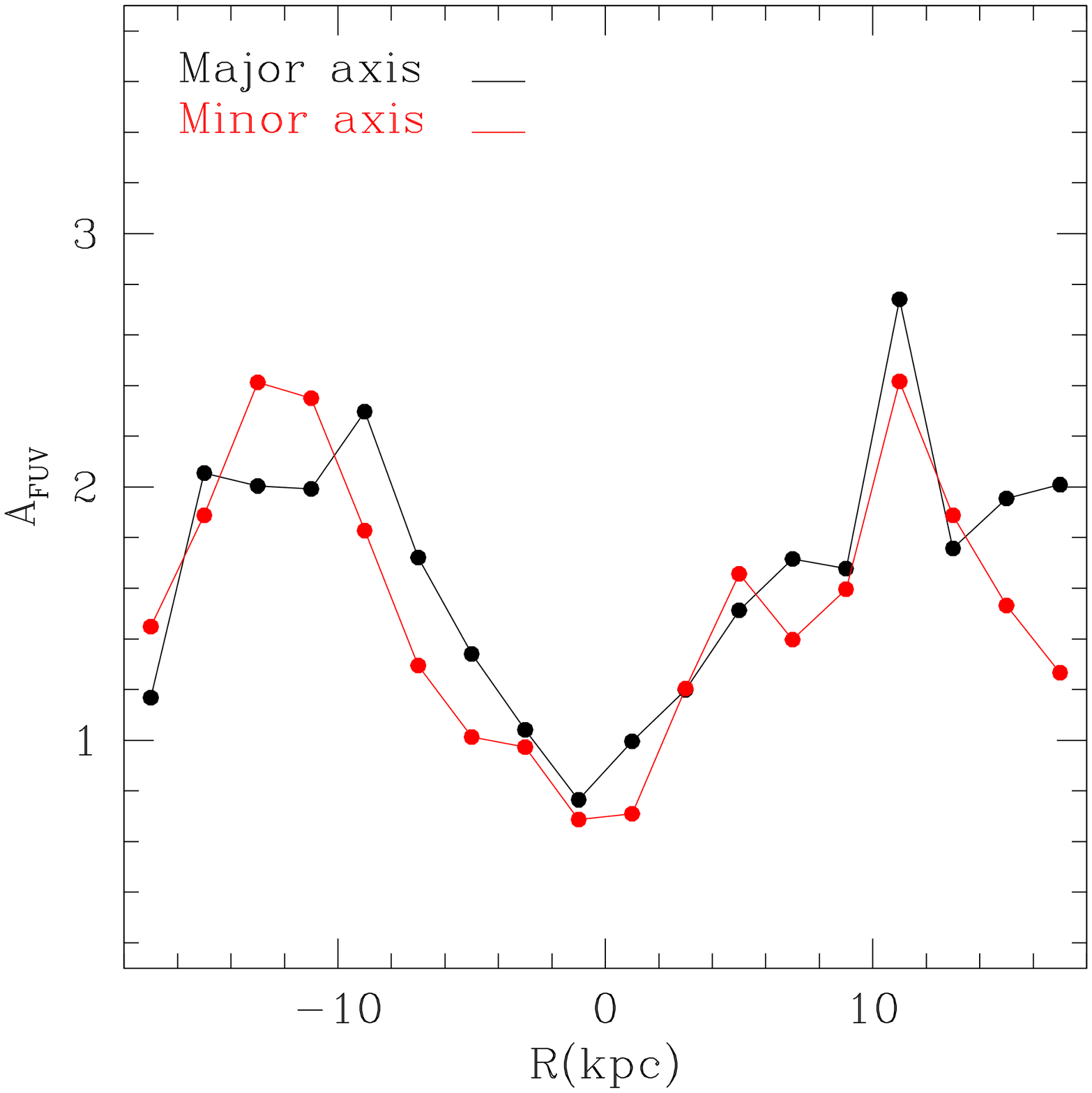}
    \caption{Left panel: $(\FUV-i_\M{SDSS})$ color profiles along
    the minor (red line) and major axis (black line). 
    Right panel: average attenuations along the minor and major axis.
    The profiles are derived considering regions with $\rm S/N>1$.
    Each point is the average value of the colors (attenuations)
    inside a 200$^{''}$-wide (0.75 kpc) stripe centered on the major
    and minor-axis in correspondence of each one of the 2kpc 
    annular regions shown in Fig.~\ref{fig:sn}}.
  \label{fig:rad_col_afuv}
\end{figure*}

  To estimate how the integrated light along the line of sight
  could have affected our analysis and consequently biased the derived
  attenuations we compared the $(\FUV-i_\M{SDSS})$ color along the
  major and minor axis as shown in Fig.~\ref{fig:rad_col_afuv}. 
  Each point in these figures is the average value of the colors (attenuations)
  inside a 200$^{''}$-wide (0.75 kpc) stripe centered on the major
  and minor-axis in correspondence of each one of the 2kpc 
  annular regions shown in Fig.~\ref{fig:sn}.
  The bias effect
  should be more evident along the minor than the major axis because
  of the inclination of \object{M31}. At the same galactocentric
  radius of the disc the line of sight passes through more central
  regions of the bulge along the minor axis than along the major
  axis. In case the described bias affects the derived parameters it
  has thus to be expected that at the same galactocentric radius the
  adopted method will predict lower attenuations along the minor axis
  than on the major axis. We calculated the mean color and attenuation
  difference inside the 10kpc ring excluding the region around +5 kpc
  on the near-side of the galaxy along the minor axis because of the
  presence of the inner spiral arm with enhanced TIR emission and
  attenuation.  In Fig.~\ref{fig:rad_col_afuv} (left panel) we see
  that indeed the color profile along the minor axis is systematically
  redder than along the major axis on average of $\sim0.5$ mag. This
  implies that the attenuations appear to be systematically lower on
  the minor axis of $A_\FUV\sim0.2$ mag (Fig.~\ref{fig:rad_col_afuv} right panel).

\begin{figure*}[!]
  \centering
  \includegraphics[width=8cm]{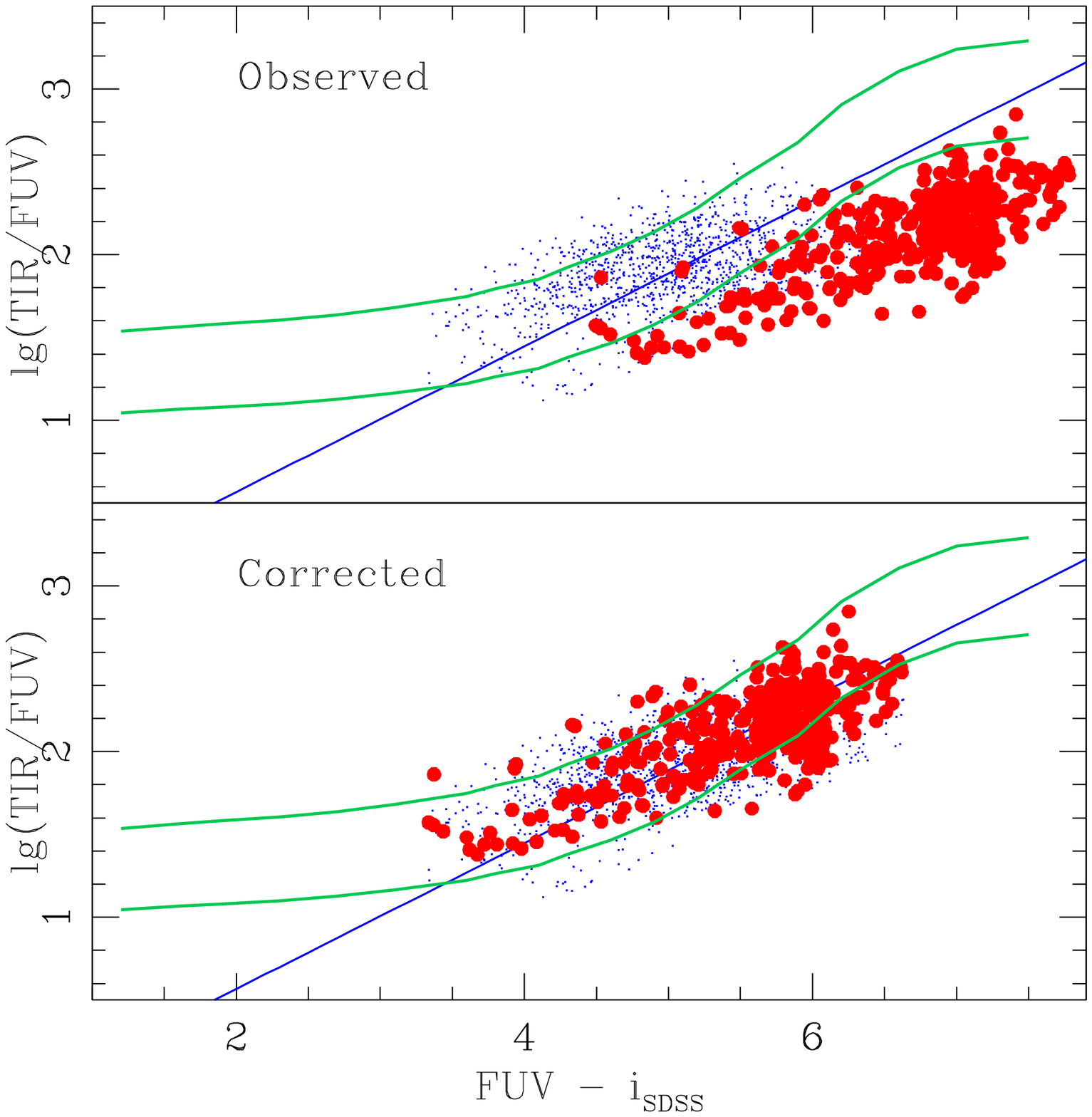}
  \caption{Upper panel: observed colors and TIR/FUV ratios for regions in the
           3kpc radial bin (big red points) and in the 11kpc radial bin (small blue points).
           Lower panel: same as upper panel, but the observed colors in the inner radial bin
           are shifted by the average color difference with respect to the best fit linear model
           in the outer ring (blue line).
    }
  \label{fig:go_TIR_FUV_comp}
\end{figure*}

  Because also the major axis should be affected by such
  observational bias the effect should be in principle larger than
  this.
Considering the radial bin at 11kpc and that one at 3kpc and considering 
the color shift toward redder values in the inner bin as $totally$ due
to the bias effect we estimated the average attenuation at 3kpc to be larger of
$\sim0.84$ mag (Fig.~\ref{fig:go_TIR_FUV_comp}) than what derived from the 
observed colors. This value should be considered as an upper limit as  
the color shift should be in part attributed also to intrinsically
redder and older populations than those in the ring where most of the star formation
in \object{M31} is taking place. Nevertheless even applying such a generous bias
correction the ring would still stick out as a maximum of attenuation as shown in 
Fig.~\ref{fig:go_TIR_FUV_comp}. On the other hand such an exercise points out the need 
of a more accurate model to properly account for what happens in the innermost
regions.

\begin{table*}[!]
  \caption{Average values and standard deviation (in parenthesis) of 
    $\lg(\TIR/\FUV)$, $(\FUV-i_\M{SDSS})$ and $A_\FUV$ in each
    annular region shown in Fig.~\ref{fig:valuemap} (upper panel). 
    The fifth and sixth columns show the linear regression
    parameters obtained fitting the data in Fig.~\ref{fig:valuemap} 
    (bottom panel) with the relation $\lg(\TIR/\FUV) = a \,(\FUV-i_\M{SDSS}) + b$.
    The last column shows the number of regions used in each annulus.
  }
  \label{tab:radial_plots} 
  \centering 
  \begin{tabular}{c c c c c c c c c} 
    \hline\hline  
    R($kpc$) & $\lg(\TIR/\FUV)$ & $(\FUV-i_\M{SDSS})$ & $A_\FUV$ & a & b & N.Reg. \\ 
    \hline 
    1. & 2.13 (0.11) & 7.51 (0.07) & 1.20 (0.14) & - & - & 9 \\
    3. & 2.15 (0.16) & 7.18 (0.45) & 1.31 (0.16) & 0.38 (0.13) & -0.55 (0.93)& 115 \\
    5. & 2.12 (0.28) & 6.68 (0.70) & 1.50 (0.30) & 0.41 (0.11) & -0.62 (0.74)& 381 \\ 
    7. & 2.05 (0.23) & 6.18 (0.75) & 1.71 (0.40) & 0.32 (0.10) & 0.10 (0.64)& 590 \\
    9. & 2.03 (0.26) & 5.66 (0.71) & 2.06 (0.44) & 0.37 (0.15) & -0.05 (0.88)& 879 \\
    11. & 1.91 (0.24) & 4.96 (0.68) & 2.45 (0.47) & 0.42 (0.24) & -0.19 (1.21)& 1200 \\
    13. & 1.81 (0.26) & 4.84 (0.69) & 2.37 (0.47) & 0.47 (0.27) & -0.45 (1.35)& 1228 \\
    15. & 1.61 (0.22) & 4.63 (0.64) & 2.17 (0.28) & 0.35 (0.10) & 0.03 (0.47)& 1380 \\
    17. & 1.59 (0.21) & 4.66 (0.80) & 2.06 (0.33) & 0.28 (0.09) & 0.29 (0.42)& 615 \\
    \hline 
  \end{tabular}
\end{table*}

  Xu \& Helou~(1996) developed a dust heating/cooling model for
  \object{M31} based on a radiative transfer code which assumes a
  sandwich geometry for stars and dust like in Cortese et
  al.~(2008). They obtained that the mean optical depth $\tau_\M{v}$
  viewed from the inclination angle of 77$^{\circ}$ increased with
  radius from $\tau_v \sim 0.7$ at $r=2\U{kpc}$, reaching a peak of
  $\tau_\M{v}=1.6$ near $10\U{kpc}$ and remaining quite flat out to
  $14\U{kpc}$. We used Eq.~2,~3,~4 of Cortese et al.~(2008) to convert
  the $\tau_\M{v}$ values of Xu \& Helou~(1996) into $A_\FUV$
  magnitudes. The resulting values are $A_\FUV = 1.23\U{mag}$ at
  $2\U{kpc}$, and $A_\FUV = 2.50\U{mag}$ near $10\U{kpc}$ in good
  agreement with the results shown in Tab.~\ref{tab:radial_plots} and
  Fig.~\ref{fig:rad_profiles}. The radial trend we derived actually
  peaks close to $10\U{kpc}$ and appears to remain quite flat in the
  outer regions and to decrease faster in the inner regions. While,
  considering the uncertainties in the $A_\FUV$ estimate, these
  results are consistent with those ones of Xu \& Helou~(1996), some
  differences are visible. In the inner side of the ring our average
  attenuations tend to be lower (from $\sim (0.1 \dots 0.5)\U{mag}$)
  with respect to the results of those authors.  Despite the fact that
  the same sandwich geometry is adopted, the two approaches actually
  differ in various aspects. At first Xu \& Helou~(1996) did not admit
  the presence of population gradients.  While this assumption was
  motivated by the consideration that the $(V-R)$ color of
  \object{M31} appeared rather constant in the disk of \object{M31}
  (Walterbos \& Kennicutt~1988) the results shown in the previous
  section, based on a much larger color baseline, indicate that indeed
  a population gradient is present across the disk of \object{M31}. In
  absence of population gradients, as discussed in the previos
  section, one tends to overestimate the attenuation in the region
  where the age effect is maximal ($4 < (\FUV-i_\M{SDSS}) < 7$), and
  this independently on the observational biases discussed previously
  (which should affect both methods in a similar way). It is
  interesting to note that in the inner region at 2kpc, and thus where
  most of the regions have $(\FUV-i_\M{SDSS}) > 7$ the two models
  appear in perfect agreement. On the other hand the model of Xu \&
  Helou~(1996) actually solves the radiative transfer problem and
  fully accounts for the effects of the dust scattering which are
  neglected by Cortese et al.~(2008). If the scatter is neglected for
  a fixed observed color we would have a smaller $\TIR/\FUV$ ratio and
  thus we would predict a smaller attenuation. While both these
  effects could have contributed to the differences we see in
  Fig.~\ref{fig:rad_profiles} the general good agreement of these
  results telling us that these effects are not dramatically changing
  our conclusions. As we have shown in the previous sections old
  populations are overall dominating the radiation field, so good
  results can be obtained if the color is fixed to the one of an old
  population as done in Xu \& Helou~(1996).  The scattering should
  affect mostly young stars embedded in the dust, and thus mainly in
  the spiral-ring structure, but as the radiation field of these stars
  is never dominant the effect should be limited.  Finally we should
  remind that we are comparing results obtained at different spatial
  resolutions ($40\arcsec/\M{px}$ in our case and $2\arcmin/\M{px}$ in
  the case of Xu \& Helou~1996) and that the accuracy of Spitzer,
  GALEX and SDSS maps is certainly larger than the measurements used
  by Xu \& Helou~(1996). In particular using IRAS images the cold dust
  emission and thus the TIR emission should be less accurately
  constrained than using Spitzer data as the longest wavelength
  measurement of IRAS is at 100$\U{\mu m}$ whereas for Spitzer is at
  160$\U{\mu m}$ close to the FIR spectrum peak of \object{M31}
  (Fig.~\ref{fig:spectrum_M31}).  In Fig.~\ref{fig:rad_profiles} we
  show also the comparison of our results with those of Hodge \& Lee
  (1998) who used two color diagrams to determine the average
  reddenings of 5 fields at different \object{M31} galactocentric
  distances. We converted the reddening values of Hodge \& Lee (1998)
  to the optical depths along the line of sight ($\tau_\M{v}$ ) by
  multiplying a factor of $2\,x\,0.921\,x\,2.8$ as in Xu \&
  Helou~(1996), then we proceeded as before to convert the
  $\tau_\M{v}$ to $A_\FUV$.  Although the radial trend always peaks at
  10kpc a larger scatter with respect to the results of Xu \&
  Helou~(1996) is visible. This may be anyway accounted by the fact
  that both us and Xu \& Helou~(1996) are analyzing radial averages
  whereas the results of Hodge \& Lee (1998) are based on 5 selected
  regions.

  The 10kpc ring is the locus where also gas density and star
  formation activity are maximum across the disk of \object{M31}
  (e.g. Braun et al.~2009; Nieten et al.~2006; Reddish~1962;
  Hodge~1979; Hodge \& Lee~1988) thus is not surprising to admit that
  also the dust density peaks at the ring as obtained also in previous
  studies.  

   Another limitation of the approach we followed
   consists in the assumption that each one of the cells we 
   analyzed is independent from the others, and thus the radiation field
   of each region is the only resposible for the dust heating in that
   region. As the mean free path of photons is typically larger than
   the dimension of the regions here anayzed ($\U{40^{''}/px}$, $\sim670\,$pc/px), it is 
   possible that photons coming from
   nearby bright regions could contribute significantly to dust heating. 
   Nevertheless this effect is likely to have a limited impact on our conclusions. 
   At first because dust heated by stars in star forming regions is 
   not the dominant source of emission in the FIR in \object{M31}, as given by the fact that
   the hot dust component emission at 24$\mu$m is overall $<7\%$ of the TIR emission as shown in
   Sec.~\ref{subsec:massestimate}. Second of all the smooth distribution of the 
   \object{M31} surface brightness in the optical ($\lambda>4000\AA$ from where most of the dust-heating 
   radiation comes) suggests the hypothesis of homogeneous conditions among adjacent regions 
   for which a local equilibrium between photon transmitted and receiveid in closeby
   regions can be assumed.

\begin{figure*}
  \centering
  \includegraphics[width=15cm]{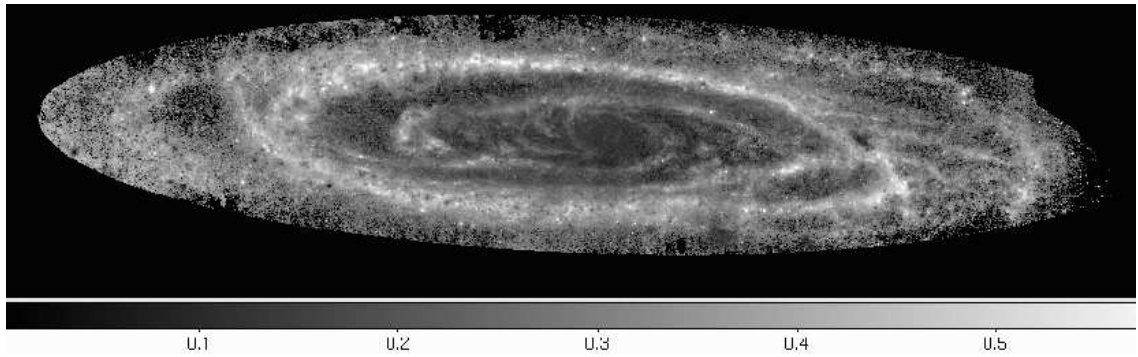}
  \caption{E(B-V) map at 6$^{''}$/px resolution ($\sim 100\U{pc/px}$ along the plane of \object{M31}).}
  \label{fig:ebv}
\end{figure*}

\begin{figure*}
  \centering
  \includegraphics[width=15cm]{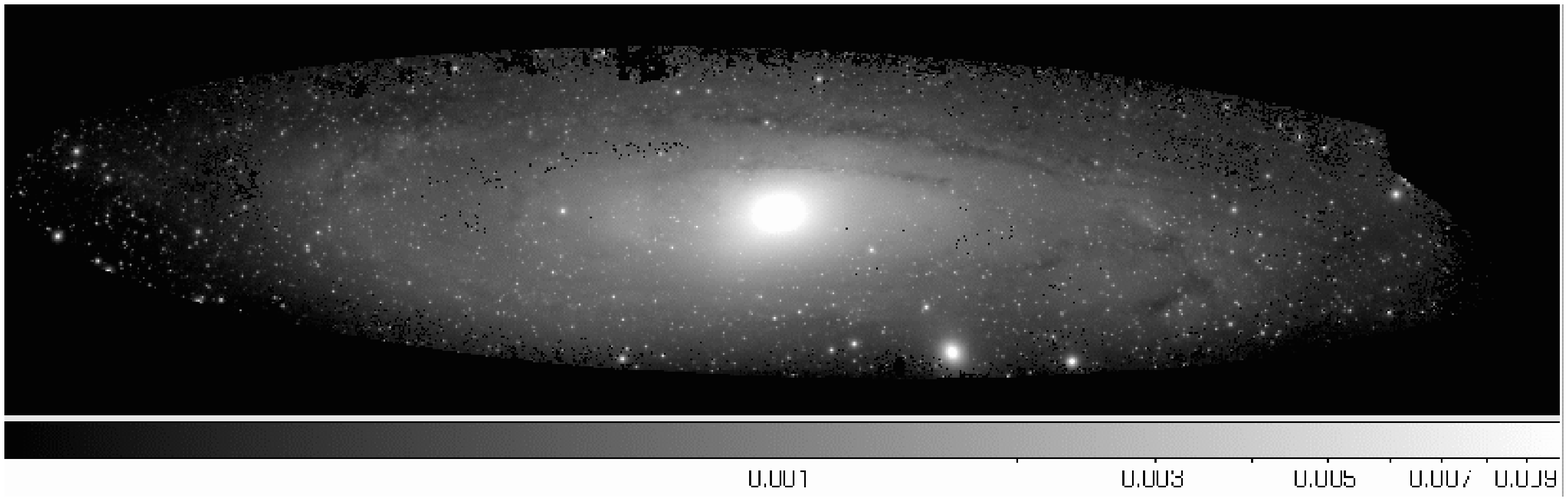}
  \includegraphics[width=15cm]{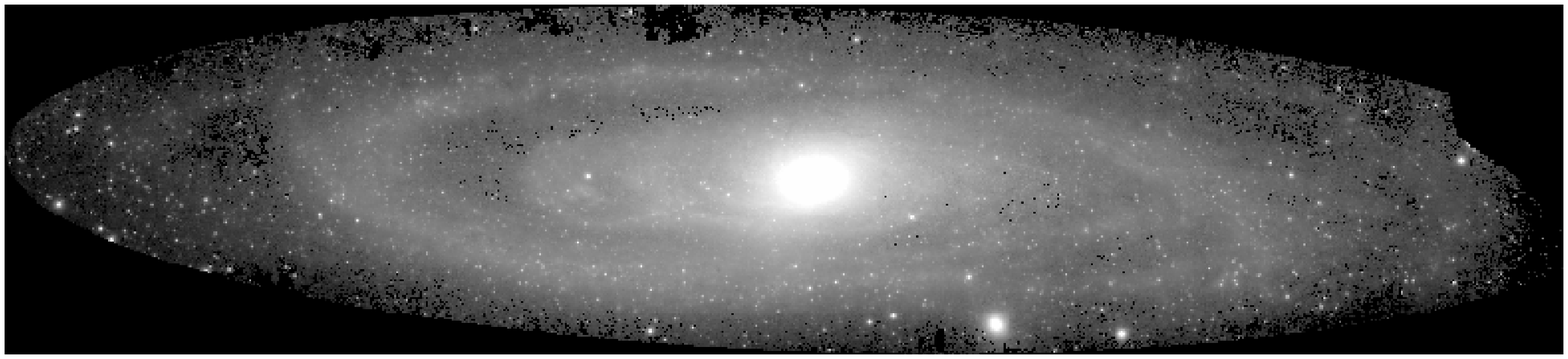}
  \caption{Upper panel: original SDSS image in the i-band. Lower
    panel: SDSS i-band image deredenned with our extinction map. For
    clarity all regions have been shown here.  Regions
    with bright point-like sources and/or with occasional negative fluxes in some
    (UV/Optical/IR) maps (black regions) are not reliable.
    The colorbar shows intensity levels in Jansky in logarithmic
    units. Images have 6$^{''}$/px resolution.  }
  \label{fig:extmap}
\end{figure*}

   Finally, using the correlation between the 24$\U{\mu m}$ and the
  TIR emission shown in Fig.~\ref{fig:corr_tir} we derived the
  reddening map of \object{M31} at 6$^{''}$/px
  (Fig.~\ref{fig:ebv}). We then applied the reddening correction to
  the observed SDSS map (Fig.~\ref{fig:extmap}). The most prominent
  dust lanes and absorption features are clearly addressed by the
  reddening map, and the ring emission at 10kpc gains importance. Note
  that in Fig.~\ref{fig:ebv},~\ref{fig:extmap} we show the whole
  maps for clarity, but regions where optical, UV and infrared
  point-like sources are visible are not reliable.  We discussed the
  issue of contaminants and how we corrected for them in
  Sec.~\ref{sec:data}.

\section{Conclusions}
\label{sec:conclusions}

In this paper we have explored some properties of the dust in
\object{M31}.  The major results we reached are the following: (i)
from the study of the infrared spectrum of \object{M31} we have
obtained that the mean intensity of the radiation field that is
heating the dust is globally low (typically $U < 2$); (ii) the dust
mass ($M_\M{dust}$) estimate remains uncertain due to the lack of
submillimetric observations, but we have obtained that $M_\M{dust}
\gtrsim 1.1\times 10^7\,M_{\odot}$, the value given by the best
fitting model being $7.6\times 10^7\,M_{\odot}$ in good agreement with
what inferred from CO and HI observations; (iii) the abundance of
Polyciclic Aromatic Hydrocarbon (PAH) particles in \object{M31} is
high ($>3\%\,M_\M{dust}$) across the spiral-ring structure of \object{M31};
(iv) we demonstrated the existence of a
correlation between the observed $\TIR/\FUV$ emission ratios and the
color $(\FUV-i_\M{SDSS})$ overall the spiral-ring structure of
\object{M31}; (vi) this correlation is not in agreement with the
$IRX-\beta$ relationship of starburst galaxies, thus color changes are
in general not driven by dust attenuations; (v) we found that 
according to the prescription of models which consider the age-dependent
dust heating the observed
correlation could be explained as the evidence of the presence of 
a population gradient and of a quite homogeneous attenuation 
of the analyzed regions; (vi) in $83\%$ of the  regions comprised
between $\rm 8kpc\,<\,r\,<\,12kpc$ the dust absorbs more than 
$50\%$ of the energy at $\lambda > 4000\,\AA$, and it appears
mainly heated by populations a few Gyr old and this could 
provide a good interpretation to the low mean
intensity of the radiation field we found from the independent
analysis of the infrared spectrum; (vii) we determined that the mean attenuation
on 2kpc-wide radial bins reaches the maximum value near 10kpc and
decreases faster in the inner than in the outer regions of the galaxy,  
and that regions with larger TIR emission have also enhanced 
attenuations; (viii) finally we derived an attenuation map of \object{M31} 
at 6$\rm ^{''}/px$ resolution ($\sim 100\U{pc/px}$ along the plane of \object{M31}).

Future contributes will investigate in more detail the dependence
of the dust attenuation values from other parameters (e.g. color,
metallicity), the star formation rate and the modellization of the
stellar populations in \object{M31}.

The maps presented in Fig.~\ref{fig:colmaps} and the E(B-V) map
of Fig.~\ref{fig:ebv} are avaiable via CDS. 

\acknowledgements{We acknowledge the useful comments of the anonymous
referee. This research was supported by the DFG cluster of 
excellence 'Origin and Structure of the Universe' 
(www.universe-cluster.de).}


\end{document}